\documentclass[a4paper]{article}
\pdfoutput=1

\usepackage{jheppub}
\usepackage{graphicx,amsmath,amssymb,todonotes,bm} 
\usepackage{caption}
\usepackage{subcaption}
\usepackage{comment}
\graphicspath{{figures/}}

\newcommand{\psb}{\bar{\psi}}
\newcommand{\AQCD}{QCD$_2$(adj)}
\newcommand*{\nc}{N}
\newcommand{\Sun}{\ensuremath{\text{SU}(\nc)}}
\newcommand{\Su}[1]{\ensuremath{\text{SU}(#1)}}
\DeclareMathOperator{\Tr}{Tr}
\DeclareMathOperator{\tr}{tr}

\newcommand{\latR}{\hat{R}}

\newcommand{\diff}{\mathrm{d}}

\newcommand*{\half}{\textstyle{ \frac{1}{2}}}
\def\Z{{\mathbb Z}}
\def\R{{\mathbb R}}

\preprint{}

\begin{document}
\title{Investigating two-dimensional adjoint QCD on the lattice}

    \author[a]{Georg Bergner}
    \author[b]{Stefano Piemonte}
    \author[c]{Mithat  \"Unsal,}

    \affiliation[a]{University of Jena, Institute for Theoretical Physics, \\
  Max-Wien-Platz 1, 07743 Jena, Germany}
  \affiliation[b]{University of Regensburg, Institute for Theoretical Physics, \\
  Universit\"{a}tsstr. 31, 93040 Regensburg, Germany}
    \affiliation[c]{Department of Physics, North Carolina State University, Raleigh, NC 27695, USA}

    \emailAdd{georg.bergner@uni-jena.de}
    \emailAdd{stefano.piemonte@ur.de}
    \emailAdd{unsal.mithat@gmail.com}

\abstract{We present our investigations of \Sun{} adjoint QCD in two dimensions with one Majorana fermion on the lattice. We determine the relevant parameter range for the simulations with Wilson fermions and present results for  Polyakov loop,  chiral condensate, and string tension. In the theory with massive fermions, all observables we checked  show  qualitative agreement between numerical lattice data and theory,  while the massless limit is more subtle since chiral and non-invertible symmetry of the continuum theory are explicitly broken by lattice regularization. 
In thermal  compactification, we observe $\nc$ perturbative vacua for the holonomy potential at high-$T$ with instanton events connecting them,  and a unique vacuum at low-$T$. At finite-$\nc$, this is a cross-over and it turns to a phase transition at large-$\nc$ thermodynamic limit.     
In circle compactification with periodic boundary conditions, we observe a unique center-symmetric minimum at any radius. 
In continuum, the instantons in the thermal case carry zero modes (for even $\nc$) and indeed, in the lattice simulations, we observe that chiral condensate is dominated by instanton centers, where zero modes are localized.  We present  lattice results  on the issue of confinement vs. screening 
in the theory and  comment on the roles of chiral symmetry and non-invertible symmetry. 
}

\date{April 2024}

\maketitle

\section{Introduction}

Perhaps, the two  most dramatic differences between the  two-dimensional  and  four-dimensional QCD with one massless adjoint Majorana fermion [QCD(adj)] are the following.  Classically, the static test charges in the two-dimensional model interact via a (confining) linear Coulomb interaction leading to a potential $V(r) \sim r$ while quantum mechanically, it is expected that $V(r) \sim (1 - \exp(-Mr))/M$, and the theory is in deconfined Higgs phase \cite{Gross:1995bp}.
  In contradistinction,  
the test charges in the four-dimensional model interact via Coulomb interaction $V(r) \sim 1/r$ while quantum mechanically, it is expected that $V(r) \sim \sigma r$, and the theory is in a confining phase \cite{Unsal:2007jx}.
  In other words, the four-dimensional theory, which is not confining classically,  becomes confining quantum mechanically and 
two-dimensional theory, which is confining classically,  becomes non-confining quantum mechanically. 

The other strange aspect of massless 2d \Sun{} QCD(adj) relative to the corresponding four-dimensional theory  is  its ground state degeneracy and relatedly its global symmetry. The standard  global symmetry in these two cases is rather similar, but the two-dimensional theory also possesses  a non-obvious global symmetry, called non-invertible symmetry \cite{Komargodski:2020mxz}.  The respective symmetries for $n_f= 1$ flavor theories are 
\begin{align} 
G=  \left\{ \begin{array}{ll}  \left[ \mathbb{Z}_N^{[1]} \rtimes (\mathbb{Z}_2)_C \right]  \times (\mathbb{Z}_2)_F\times (\mathbb{Z}_2)_\chi    \times G_{\rm non-inv} & \qquad {\rm  2d \; massless\; QCD(adj) } \cr
  \left[ \mathbb{Z}_N^{[1]} \rtimes (\mathbb{Z}_2)_C \right]  \times (\mathbb{Z}_2)_F\times (\mathbb{Z}_{2\nc})_\chi    & 
   \qquad {\rm  4d  \; massless\; QCD(adj) } 
\end{array}  \right. \; ,
\end{align}  
while the respective number of vacua are given by 
\begin{align} 
{\cal N}(\rm vacua)=  \left\{ \begin{array}{ll}  2^{\nc-1}  & \qquad {\rm  2d \; massless\; QCD(adj) } \cr
 \nc   &   \qquad {\rm  4d \; massless\; QCD(adj) } 
\end{array}  \right.\; .
\end{align} 
While the slight change in the discrete chiral symmetry is due to nature of Majorana fermion in 2d vs.\ 4d, 
the exotic vacuum degeneracy is  dictated by representations of  the non-invertible symmetry.\footnote{Furthermore, to add more  to the strangeness of 2d theory, these $2^{\nc-1}$ vacua are split to $N$ universes, sectors which are separated by \emph{ non-dynamical} domain-walls, charged under 1-form symmetry. Each universe supports about $2^{\nc-1}/\nc$ vacua separated by standard dynamical domain walls. We review this structure in Sec.~\ref{sec:univ}. }

These two facts should make it clear that the goal of studying QCD(adj) in two dimensions is not  necessarily to have a simpler toy model for the dynamics of a strongly coupled four-dimensional theory. Rather, one sensible goal is to gain a deeper and broader understanding of quantum gauge theories and their possible behaviors.  

In two-dimensional pure Yang-Mills theory, there are effectively no dynamical gauge degrees of freedom,  and  both classical and quantum theory exhibit linear confinement, as it can be shown by analytical methods \cite{Migdal:1975zg}. 
The theories become non-trivial as soon as bosonic or fermionic matter fields are added.  
The situation with \Sun{} QCD(adj), or Yang-Mills with one Majorana fermion is rather special as described above.     
As explained, the  exactly massless theory with standard fermion term $\tr [\psb \gamma_\mu D_\mu  \psi ]$  is non-confining, 
a surprising behaviour as pointed out in  \cite{Gross:1995bp}. For a more general perspective, we have to consider additional  deformations by other operators.
Once a chiral symmetry breaking mass term is turned on, the theory becomes confining. 
However, about twenty-five years after this first analysis, it has been understood that two-dimensional QCD(adj) (\AQCD) admits certain classically marginal four-fermion deformations \cite{Cherman:2019hbq}, just as the Gross-Neveu model \cite{Gross:1974jv},   which can become relevant in the quantum theory \cite{Cherman:2024onj}. The result of \cite{Cherman:2019hbq} is based on  mixed anomalies between the standard symmetries of the system, as well as the notion of adiabatic continuity and  semi-classical limit in exact agreement with anomalies.   These deformations make the massless theory confining for all representations (except for  $N$-ality $\nc/2$ for even $N$), proving that the mass term is not the only deformation inducing confinement in this theory.  

Motivated by the combination of the results of \cite{Gross:1995bp} and 
\cite{Cherman:2019hbq}, Ref.~\cite{Komargodski:2020mxz} made a further significant progress. It showed that the massless \AQCD{} with just standard fermion terms $\tr [\psb \gamma_\mu D_\mu  \psi ]$ 
has an extra global symmetry, called  non-invertible symmetry, $G_{\rm non-inv}$. The practical utility of this symmetry is that  it forbids the four-fermion deformations which are allowed by standard  global symmetries.  

The four-fermion operators  are meant to be for the non-invertible symmetry what the mass operator  is  for 
chiral symmetry.  
These operators  are charged under the respective symmetry, but we are free to study the behaviour 
of the theory by turning  them on.   Indeed, Ref.~\cite{Komargodski:2020mxz} studied the response of the theory by turning on four-fermion deformations and reached to the same conclusion as \cite{Cherman:2019hbq}, that all  representations except for  $N$-ality $0,\nc/2$ exhibit confinement. This result  actually tells us that \cite{Gross:1995bp} and  \cite{Cherman:2019hbq} are not conflicting with each other.  Some other work on \AQCD{} can be found in \cite{Smilga:1994hc,Smilga:1996dn,Cherman:2019hbq,Delmastro:2021xox,Komargodski:2020mxz, Kutasov:1993gq, Kutasov:1994xq, Dempsey:2022uie, Trittmann:2023dar,Dempsey:2023fvm,Cherman:2024onj, Katz:2013qua, Lenz:1994du}.

We now understand that an  analogous,  but relatively simpler story holds in charge-$q$ version  \cite{Anber:2018jdf, Anber:2018xek, Armoni:2018bga, Misumi:2019dwq, Honda:2021ovk} 
of   Schwinger model~\cite{Schwinger:1962tp}.  
  It is usually asserted that  Schwinger model with massless fermions always exhibits screening, while its massive deformation confines. 
  These statements have a precise meaning in the charge-$q$ version which possess a  $\Z_{q}^{[1]}$ 1-form symmetry. 
  However, a four-fermion deformation of the Schwinger model, which respects a $\Z_{4}$ subgroup of $\Z_{2q}$ chiral symmetry, which is large enough to prohibit a mass term for $q={\rm even}$,  is shown to be confining if the four-fermion deformation is relevant \cite{Cherman:2022ecu}, mimicking the story in \AQCD.    

The present work is an initial study to explore some of the interesting dynamics of  \AQCD{}   
by using lattice gauge theory, a  manifestly non-perturbative framework. 
For the evaluation of finite temperature chiral condensate and Polyakov loop properties, we use Wilson fermions. To evaluate the zero temperature chiral condensate, we use overlap fermions by reweighting. 
Lattice formulation  proved to be very useful  for the investigation of QCD and Yang-Mills theories in $\R^4$ by simulating the theories on   large discretized torus $T^4$.  
In two dimensions these well-established methods also provide important cross-checks and insights. 
The lattice studies of two-dimensional gauge theories have so far mainly considered the Schwinger and 't Hooft models \cite{Berruto:1999cy,Berruto:2002gn}. Some very preliminary studies exist also for the case of adjoint QCD in two dimensions \cite{Korcyl:2011kt}. Recently, an interesting alternative approach using a lattice Hamiltonian formulation has been applied in the case of SU(2) \cite{Dempsey:2023fvm}. Some  related  supersymmetric gauge theories have been already investigated on the lattice in some numerical studies \cite{Kanamori:2008yy,August:2018esp}.
In addition, two-dimensional gauge theories have been an ideal playground to test numerical methods, algorithms, and approaches. 

The main results of our numerical simulations, which always approach the theory from a finite mass regime, are the following. In thermal case, 
absolute value of the Polyakov loop makes a cross-over from 
small $|\tr P|$ towards $|\frac{1}{N}\tr P| \sim 1$ at some $T_c \sim g$. At high-$T$, we find $N$ vacua and tunnelings (instanton events) connecting them. Due to tunnelings, $\langle \tr P \rangle =0$ in both phases and strictly speaking, there is no phase transition at finite-$N$. In circle compactification with periodic boundary conditions, we observe a unique vacuum for Polyakov loop, i.e.\ small $|\tr P|$ even at small circle with global minimum of the effective potential around $\tr P=0$. We determine a chiral condensate  at low and high temperature. The high-temperature condensate is dominated by the instanton events interpolating between the $N$ vacua. Concerning confinement vs. screening, at finite values of $m$, we observe a finite string tension. 
In the chiral limit, the tension tends to zero consistent with screening. Note, however, that our simulations may not be precise enough to capture the effects of four-fermion induced confinement.  

The paper is organized as follows: In Sec.~\ref{sec:overview}, we first overview briefly what is known about the continuum theory 
on $\mathbb R^2$ and  $\mathbb R \times S^1$, both in thermal  and  circle compactification suitable for adiabatic continuity.
 We apply quite standard techniques for the simulations of the theory and therefore only a short discussions of the methods is presented in Sec.~\ref{sec:lat}. Since the theory is distinguished from four dimensional counterparts, we provide a longer discussion of the relevant parameter range and tuning in Sec.~\ref{sec:parameter_tuning}. In Sec.~\ref{sec:obs},  we present  data for the observables of the theory, showing the zero temperature and finite temperature fermion condensates,  Polyakov loop, its modulus and susceptibility  as observables. In order to provide a careful study of the confinement properties, we present different methods to estimate the string tension in Sec.~\ref{sec:string_tension}. Finally we discuss some preliminary results on the lightest boson and fermion masses.
 
\section{ Overview of continuum theory on $\mathbb R^2$ }
\label{sec:overview}
We are considering SU($\nc$) Yang-Mills theory in two dimensions coupled to an adjoint Majorana fermion (\AQCD).
The Lagrangian of this theory is\footnote{
The conventions for the Euclidean $\gamma$ matrices 
$\gamma_0=\sigma_1$;
$\gamma_1=-\sigma_2$, which means
$\gamma_\ast=\gamma_3=i\gamma_0 \gamma_1=\sigma_3$.
The charge conjugation matrix is $C=\gamma_1$. $0$ direction corresponds to $x$ (length $\widetilde L=L_x$), $1$ direction corresponds to $t$ (length $L=L_t=1/T$). $t$ is the direction of thermal/periodic compactification.}
\begin{align}
 S= \int \diff^2 x  \frac{1}{2g^2}  \tr [F_{\mu\nu}F_{\mu\nu}] + \tr [\psb\left(\gamma_\mu D_\mu  \right)\psi ] \; ,
\label{lag}
\end{align}
with gauge field strength $F^a_{\mu\nu}$ and a Majorana fermion $\psi$ in the adjoint representation. Instead of the gauge coupling $g$, we can also use the 't Hooft  coupling $\lambda=g^2 \nc$. $D_\mu$ is  
the covariant derivative. 

Massless \AQCD\ possesses ordinary 0-form, 1-form, and non-invertible  symmetries, which makes 
their interplay quite interesting. As explained, the complete symmetry group is in this case
\begin{align} 
G=  \left[ \mathbb{Z}_N^{[1]} \rtimes (\mathbb{Z}_2)_C \right]  \times (\mathbb{Z}_2)_F\times (\mathbb{Z}_2)_\chi    \times G_{\rm non-inv}\, .
\label{sym}
\end{align} 
The theory admits a relevant mass deformation and two marginally relevant four-fermion deformations:\footnote{The mapping to chiral basis is as follows: 
$ {\cal O}_{\chi} = 
\tr[\psi_{+}\psi_{-}],  \; 
{\cal O}_1 =\tr(\psi_{+}\psi_{+} \psi_{-} \psi_{-}) ,  \;  {\cal O}_2 
= \tr(\psi_{+} \psi_{-})\tr(\psi_{+} \psi_{-})
$.  
}  
\begin{align}
&m {\cal O}_{\chi} = 
m \; \tr[  \bar \psi  \psi ]  \qquad  &  { \rm breaks } &\;  (\mathbb{Z}_2)_\chi    \times  G_{\rm non-inv}    \nonumber \\
&c_1 {\cal O}_1  + c_2 {\cal O}_2 = 
 c_1 \tr[ \bar \psi \gamma_{\mu}\psi 
\bar \psi \gamma_{\mu}\psi ] + c_2 \tr[ \bar \psi \psi ] \tr  
[\bar \psi \psi],   
\qquad  &  { \rm breaks }&  \; G_{\rm non-inv}  \, .
\label{eq:def}
\end{align}
Since the canonical dimension of fermions in two dimensions is $[\psi]=1/2$, it is natural  to consider these four-fermion deformations as proposed in  \cite{Cherman:2019hbq}.
These operators are classically marginal, i.e.\  their coefficients are dimensionless. These four-fermion operators respect all standard (invertible) symmetries, but ${\cal O}_2$ violates the non-invertible symmetry.
One can show that the  coupling constants 
are  asymptotically free for a certain choice of the couplings,  \cite{Cherman:2024onj},  just like in the Gross-Neveu model \cite{Gross:1974jv},  hence the operator is marginally relevant and alters the IR dynamics. 

It has been shown in Ref.~\cite{Komargodski:2020mxz} that non-invertible symmetry $G_{\rm non-inv}$ forbids the dangerous one of these four fermion operators, ${\cal O}_2$,  just like chiral symmetry   $(\mathbb{Z}_2)_\chi  $ forbids the mass operator. 
Therefore, in a microscopic regularization which respects 
$G_{\rm non-inv}$, these operators  are not be generated and one can set their coefficients consistently to zero. However, if regularization does not respect $G_{\rm non-inv}$,  these operators  are generically generated, and their effects can alter the IR dynamics.

\begin{figure}[t]
\begin{center}
\includegraphics[angle=0, width=0.6\textwidth]{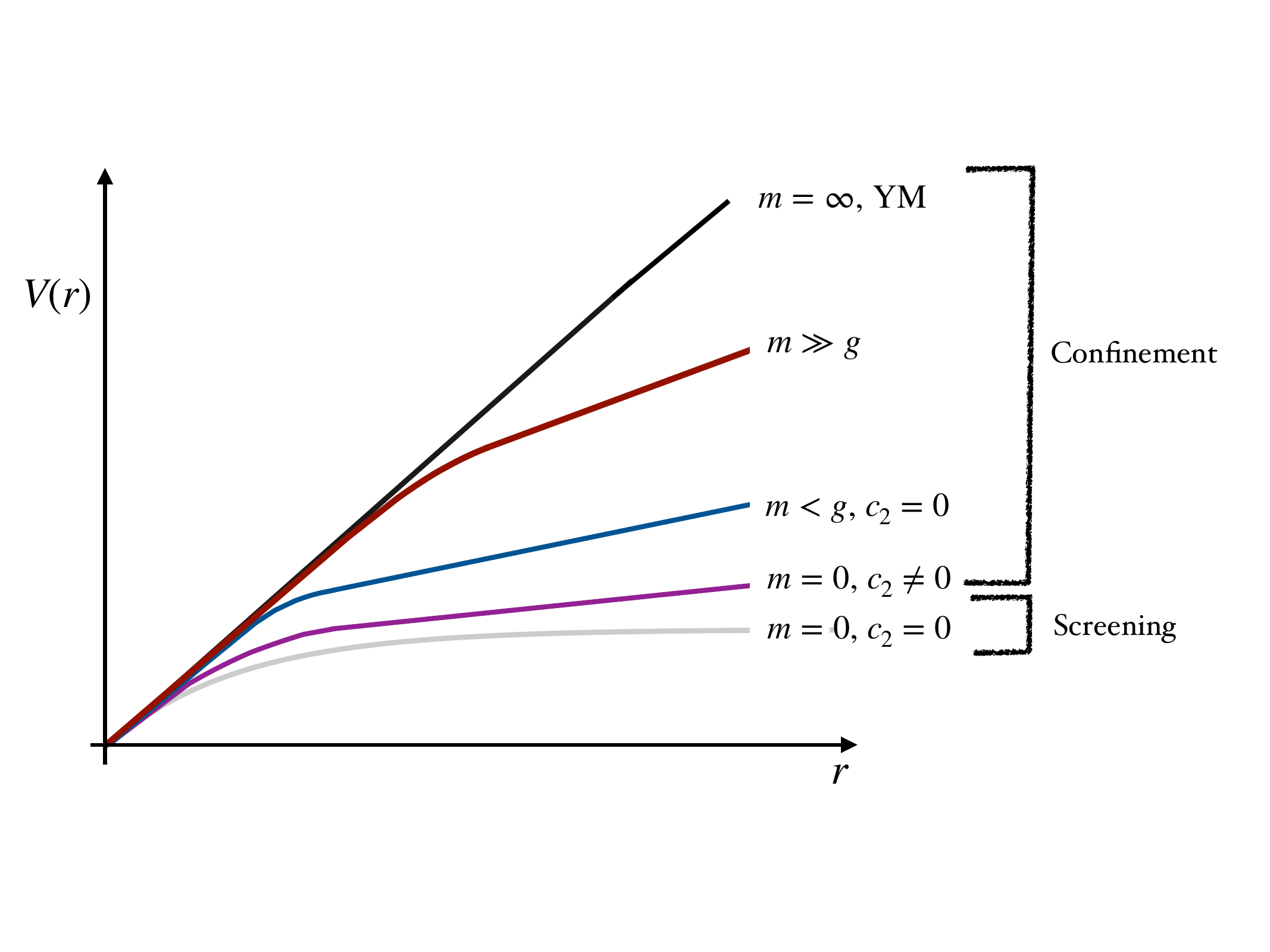}
\vspace{-1.5cm}
\caption{Cartoon of potential between test charges in a generic representation in $SU(N)$ QCD(adj) $N \geq 3$. The theory is in the screening phase if  $ (\mathbb{Z}_2)_\chi    \times G_{\rm non-inv}   $ breaking perturbations  are turned off,  $m=0, c_2=0$.  
Turning on 
 either perturbations leads to confinement except for the $N$-ality $N/2$ in the case of $c_2 {\cal O}_2$ deformation. The relatively new realization is that exactly massless theory becomes confining with the $c_2$ deformation as shown by using mixed anomalies  
 and explicit soft breaking of  non-invertible  symmetry. At infinite mass limit, the tensions are representation dependent rather than $N$-ality.   The change in slope for $m \gg g$ is a transition from a representation dependent tension to $N$-ality dependent tension.  
  }
\label{Potential}
\end{center}
\end{figure}

\noindent
\subsection{Confinement vs.\ screening} 
\label{sec:con-sc}
The feature that makes \AQCD{} a rather interesting theory is the fact that the explicit breaking of  either of the 
$ (\mathbb{Z}_2)_\chi $ and/or $ G_{\rm non-inv}   $ symmetries  changes the dynamics drastically. 
Note that an 
 $ (\mathbb{Z}_2)_\chi $ breaking perturbation also  necessarily  breaks $G_{\rm non-inv}$. However, the inverse is not true. The four-fermion deformation considered above breaks $G_{\rm non-inv}   $, but  
  respects   $ (\mathbb{Z}_2)_\chi $.   The resulting outcomes on  $\mathbb R^2$ are the following: 
\begin{itemize} 
\item Classically, the theory with   $ (\mathbb{Z}_2)_\chi \times G_{\rm non-inv}$ symmetry $(m=0, c_2=0)$    has a linear Coulomb potential among test charges, but quantum mechanically,  it is deconfined.  
 The quantum theory has an exponentially large number of vacua,   $2^{\nc-1}$, half bosonic and half fermionic. 

\item A mass deformation of the theory, $m \neq 0$,  explicitly breaking both   $ (\mathbb{Z}_2)_\chi    \times G_{\rm non-inv}   $,  leads to confinement, with a string tension $\sigma \sim m g$ at long distances. The vacuum is expected to be unique.  

\item A four-fermion deformation which explicitly breaks  $  G_{\rm non-inv}   $ but respects  $ (\mathbb{Z}_2)_\chi $  is also expected to lead to confinement, similar to four-fermion deformation of the massless Schwinger model \cite{Cherman:2022ecu}. The number of vacua is 2 for $\nc \neq 4n +1$ ($n=1,2,\ldots$), and otherwise there is a unique vacuum.

\item If we explicitly break  the non-invertible symmetry $G_{\rm non-inv}$, and keep 
the ordinary global symmetries \eqref{sym} intact, then the massless theory confines all representations except for $N$-ality $0, N/2$, i.e.\ 
\begin{align} 
\sigma_k  \neq 0  \quad {\rm for} \;  k \neq 0, N/2,   \quad  
\text{and symmetry breaking}\quad\Z_N \rightarrow \Z_{N/2}\; .
\end{align}

\end{itemize} 

In the $m \rightarrow \infty$ limit, \AQCD{} reduces to pure Yang-Mills theory without fermions. This is an almost topological theory without any physical degrees of freedom.  This theory can be solved exactly.
The string tensions can  be  computed exactly \cite{Migdal:1975zg}. In the case of SU(2) gauge theory, the string tension in representation $j=1/2,1,\ldots $ is 
\begin{align}
    \sigma_j(\beta)=
    -\log\left( \frac{I_{2j+1}(\beta)}{I_1(\beta)}\right)  
    \xrightarrow
    [{\beta \rightarrow \infty}]{}  \frac{1}{\beta} j (j+1)  \equiv \frac{1}{\beta} C_2(j)\, ,
    \label{rep}
\end{align}
where $I_n(x)$ are modified Bessel functions. See Section 14 of Ref.~\cite{Wipf:2021mns} for details. The fact that string tensions are not classified under the $\Z_2$ center of SU$(2)$, but rather in terms of quadratic Casimir of the spin-$j$ representations is explained in terms of another non-invertible symmetry. This non-invertible symmetry is a 1-form symmetry $G^{[1]}_{\rm non-inv}$, which is distinct from the previously discussed 0-form non-invertible symmetry $G_{\rm non-inv}$ in \AQCD{} \cite{Nguyen:2021naa, Pantev:2023dim}.     $G_{\rm non-inv}^{[1]}$ explains why we have infinitely many types of string tensions rather than just $N$-types from symmetry perspective. 
Of course, dynamically, this is obvious as the pure Yang-Mills theory does not have charged gluons  
to screen the adjoint probe charges. 
 The 1-form non-invertible symmetry $G^{[1]}_{\rm non-inv}$ in pure Yang-Mills is explicitly broken by the introduction of the matter field in adjoint representation down to $\Z_N^{[1]}$  1-form symmetry.  It is an interesting problem how the confining ${\cal R}$ dependent  strings  of pure Yang-Mills theory 
become confining $k=|{\cal R}|$ $N$-ality dependent  strings in massive QCD(adj).

We can summarize our knowledge about string tension  for general \Sun{} theory as follows. Let ${\cal R}$ denote a representation of \Sun{} and   $ k= |{\cal R}|$ the $N$-ality of the representation. Then, the string tensions take the form:
\begin{align} 
\sigma_{\cal R}  &= 0 \qquad \qquad \qquad \qquad  \;\; \; \;   m =0  \qquad \qquad     \Z_N^{[1]}  \;\;   {\rm broken}   \cr 
\sigma_{\cal R}  & =  m \lambda^{1/2} \sin \left( \frac{ \pi k}{\nc} \right), \qquad   
m \ll \lambda^{1/2}  \qquad   {\Z_N^{[1]} \;\;   \rm unbroken}  \cr
\sigma_{\cal R} & =  \lambda C_{\cal R} \qquad \qquad \qquad \qquad   m \rightarrow \infty \qquad   \;\;  {G^{[1]}_{\rm non-inv} \;\;   \rm unbroken} , 
\label{current}
\end{align}
where  $C_{\cal R}$ is quadratic Casimir for representation ${\cal R}$.   These theoretical expectations are sketched in Fig.~\ref{Potential}. 
Our simulations may be viewed as   probing the properties of the mass-deformed theory, including an extrapolation 
 towards the chiral (i.e.\ massless) limit.

\subsection{Vacuum  degeneracy, universes vs. ordinary vacua}
\label{sec:univ}
We add a few remarks on the vacuum degeneracy of the massless theory.  As stated above, the theory has $2^{N-1}$ fold vacuum degeneracy.    However, some peculiarities of these vacua require understanding the distinction between super-selection sectors and a stronger notion, called the  ``universes".   

For ordinary discrete symmetry breaking in  two Euclidean dimensions, 
domain walls are finite tension configurations connecting the vacua.  Consider  a 1+1 field theory with a $\Z_2$ symmetry,  and a potential 
$V(\phi) = \half (\phi^2 - v^2)^2$.  In the broken regime, there are two degenerate vacua on $\R^2$, and there exists a field configuration  (called a kink) 
at the interface of the two vacua. A kink can be viewed as a particle in the spectrum of the theory. Now,  consider the charge-2   $(q=2)$ Schwinger model. This model has a  $(\Z_2)_{\chi} \times \Z_2^{[1]}$ chiral and 1-form symmetry, with a mixed anomaly in between.  Let ${\cal O}_{\chi} (x)  \equiv \bar \psi_{\rm R}  \psi_{\rm L}(x)  $ denote  the chiral fermion bilinear. 
  The mixed  anomaly implies that 
  \begin{align}
  {\cal O}_{\chi} (x)   W(C) =  e^{ 2 \pi i \;  {\rm Link} (C, x) /2} W(C)  {\cal O}_{\chi} (x) \; ,
  \label{ano-s}
  \end{align}
   i.e.\   
  ${\cal O}_{\chi} (x) $ acts as the local topological operator generating the 1-form symmetry, and it measures the charge associated with the Wilson line. Here, ${\rm Link} (C, x)$ is the linking number of $x$ and  $C$.  
  Now, assume ${\cal O}_{\chi} (x)   | 0 \rangle = + | 0 \rangle$ in one of the vacua.  Acting on the 
$|0 \rangle$  with a space-like infinite Wilson line $W$ and using \eqref{ano-s}, one can show that the state $W|0 \rangle \equiv |W \rangle $  obeys  ${\cal O}_{\chi} (x)  |W \rangle = {\cal O}_{\chi} (x) W|0 \rangle = - W {\cal O}_{\chi}  |0 \rangle = 
-|W \rangle$. This is  other chirally broken state.  Therefore, $W(C)$ is the boundary between the two chirally broken vacua.  
This makes  the difference compared to ordinary $\Z_2$ breaking (not involving anomalies) manifest.  In the former, the kink is  a particle in the spectrum of the theory, a finite tension dynamical domain wall.  However, in the second model,  the ``kink"   has a charge, $+1$ under the $\Z_2^{[1]}$ 1-form symmetry. It is non-dynamical and can only be introduced to the theory as an external probe. In other words, the domain wall in the second  theory has infinite tension.  
In particular,  in the first model, if the space is compactified on a circle, $\R_{t} \times S^1_L$, space-independent saddles  $\phi(t)$  represent tunneling and the true ground state is the symmetric superposition, $ \frac{1}{\sqrt 2}  (  |-a\rangle  +   |a\rangle ) $ separated from anti-symmetric combination by $e^{- M L}$, where M is domain-wall mass.  In the second model, 
 no tunneling exists between the broken vacua even if the theory is compactified.  The superselection sectors separated by these stronger criteria are called universes.  

\AQCD{} has a  $\Z_N^{[1]}$ 1-form symmetry,  for which the generators are  local  topological operators 
${\mathsf U_{\mathsf s}} (x)$.\footnote{Definition of ${\mathsf U_{\mathsf s}} (x)$: delete the point $x$ from the spacetime, and perform the path integral over gauge fields with holonomy  ${\rm hol}_C (x) = e^{2 \pi i n/N} {\bf 1}, n=0,1, \ldots, N-1$ for small clockwise circles circulating $x$. The insertion of the  operator ${\mathsf U_{\mathsf s}} (x)$  is equivalent to
performing the path integral with non-trivial 't Hooft magnetic flux sector $n \in \Z_N$.  }      The implication of  $G_{\rm non-inv}$, on the other hand, is that there exist $O(2^{2N})$ {\it topological line operators}  $\alpha(C)$. 
These two types of operators  satisfy 
\begin{align} 
{\mathsf U_{\mathsf s}} (x) \alpha(C) = e^{2 \pi i  /N} \alpha(C) {\mathsf U_{\mathsf s}} (x) \; .
\label{algebra}
\end{align}
   If we act on the vacuum  $|0 \rangle$ with a topological line, then the state $\alpha|0 \rangle$  has a different   charge  under the 1-form symmetry relative to  the original vacuum, i.e.\ the domain wall connecting the two vacua is charged under $\Z_N^{[1]}$.   
    However, in the \AQCD, there are no fundamentally charged matter fields. Hence, we have to view the vacua with different  1-form symmetry charge as distinct universes. 
    
    Physically, these 
   universes can be thought of as the theories on the sector 
    generated from the vacuum by different $N$-ality probes, for example 
    charges $\pm 1$ at $ \mp \infty$, respectively.    
   The algebra 
  \eqref{algebra} implies  that  there are  $N$ different types of universes distinguished by their charges under the 1-form symmetry. 
  Since there are $2^{N-1}$ vacua in the theory, 
 each universe must support multiple degenerate vacua. 
   The number of vacua in each universe   ${\rm dim}[{\cal U}_{N,k}]$ is  roughly $2^{N-1}/N $.  For example, 
 for $N=3,4,5,6,9$,  the number of vacua in each universe is given by:
 \begin{align}  
 \label{eq:universe_vacua}
  N=3 \qquad  &{\rm dim} [{\cal U}_{3,k}] = 2,1,1 \qquad  &&{\rm for} \; k=0, 1, 2 \cr
    N=4 \qquad   &{\rm dim} [{\cal U}_{4,k}] = 2,2,2,2 \qquad && {\rm for} \; k=0, 1, 2,3 \cr
 N=5  \qquad  &{\rm dim} [{\cal U}_{5,k}] = 4,3,3,3,3 \qquad&& {\rm for} \; k=0, 1, 2, 3, 4, \cr 
 N=6  \qquad  &{\rm dim} [{\cal U}_{6,k}] = 6,5,5,6,5,5 \qquad && {\rm for} \; k=0, 1, 2, 3, 4,5  \cr
 N=9  \qquad  &{\rm dim} [{\cal U}_{9,k}] =  30, 28, 28, 29, 28, 28, 29, 28, 28\qquad && {\rm for} \; k=0,  \ldots, 8 
 \end{align}

\noindent
{\bf Does lattice formulation possess a microscopic non-invertible symmetry? }
In our work,  we mostly use a lattice formulation with Wilson fermions,  neither chiral symmetry, nor 
$G_{\rm non-inv}$  are respected. In particular, the Wilson term,  $r a \psb  D^2 \psi$, in loops generates an $O(a^0)$  additive mass renormalization for fermions, which introduces the ${\cal O}_{\chi}$ operator.  It also generates, via  tree-level gluon-exchange diagram, a four-fermion operator with a dimensionless effective coupling $ c_2 \sim r^2 a^2 g^2$ \cite{Cherman:2024onj}, the  ${\cal O}_2$ operator. ${\cal O}_1$ will also be generated because it is not protected by any symmetry.    Therefore, the lattice \AQCD{} in 2d, is in general a multi-scale problem. In particular,  the scale associated with coupling $g^2$, and  the strong scale $\Lambda_{c_2}$  associated with  $c_2$ as implied by asymptotic freedom should be particularly important.   
Within the range of our simulations,  we do not see the effects of the induced
four-fermion terms,  very likely because of  the volumes we work with,  i.e., as if we have accidental non-invertible symmetry in the IR.  Perhaps, due to the logarithmic running of $c_2$, $\Lambda_{c_2}^{-1}$ is bigger than the box size, and we do not probe its effects yet. 
It may be useful  to tune the coefficient of these operators as desired from the beginning (for example, to strong coupling)  to see the effect of these operators on  the dynamics of the theory. We leave this as an interesting open problem. 

\subsection{On $\mathbb R \times S^1$, with  periodic and anti-periodic boundary conditions} 
\label{sec:apbc}
In order to  understand some aspects of center-symmetry in \AQCD{} on $\R \times S^1$, we compactify  the theory on a circle and study its  gauge holonomy potential.    
In the simulations, we consider both  anti-periodic  and periodic   boundary conditions for fermions, where the former is thermal compactification corresponding to partition function    and the latter is circle compactification which corresponds to a 
$(-1)^F$ graded partition function: 
\begin{align} 
 Z(L) &= \tr [e^{-L H}],  \qquad   \qquad {\rm thermal \;  comp.}  \cr
 \widetilde Z(L) &= \tr [e^{-L H} (-1)^F]   \qquad {\rm circle \;   comp.} 
 \end{align} 
 In the thermal case, $L$ corresponds to an inverse temperature $1/T$. For periodic compactification, $L$ is not related to temperature.  
 Let us denote the Polyakov loop 
going around the circle by $P$.  Up to gauge conjugations, it can be written as:  
\begin{align}
P = {\rm diag} \left( e^{i \theta_1}, e^{i \theta_2}, \ldots,  e^{i \theta_N}  \right),  \qquad \sum_{i=1}^{N}  \theta_i = 0
\end{align}
 The holonomy potential can be reliably computed at weak coupling, i.e, at sufficiently small circle 
 $g^2 L^2 \ll1$.  
  Since the theory does not have bosonic propagating degrees of freedom, the holonomy potential is completely dictated by adjoint fermions. 
  
\begin{figure}[t]
\begin{center}
\includegraphics[angle=0, width=0.8\textwidth]{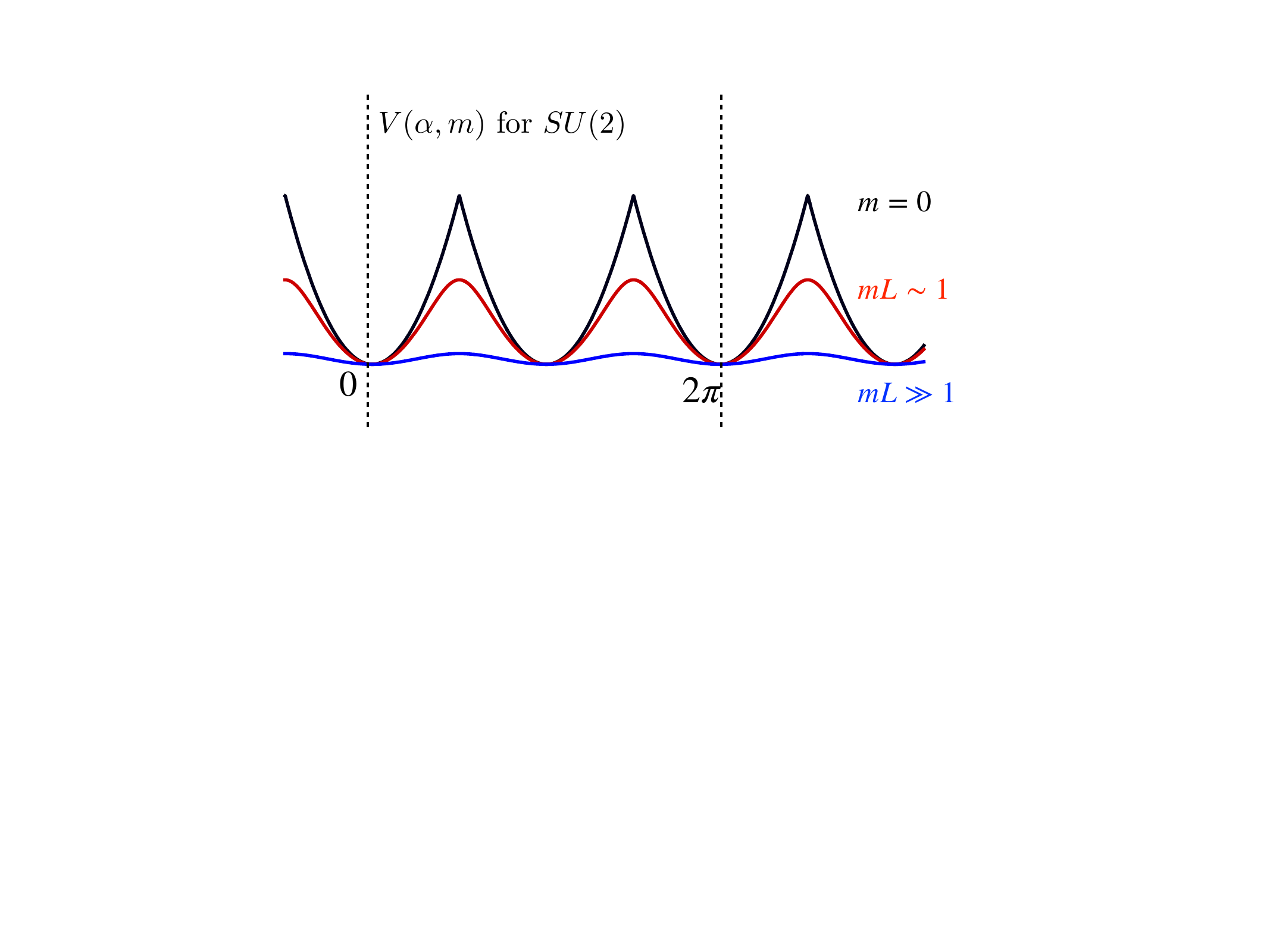}
\vspace{-5.2cm}
\caption{Holonomy potential for $SU(2)$ gauge theory as a function of fermion mass. 
With increasing  $mL$, the potential flattens.  At  $m=\infty$, pure Yang-Mills theory in 2d, holonomy potential is zero.  
 }
\label{Hol-m}
\end{center}
\end{figure}

For periodic $(+)$ and anti-periodic $(-)$ boundary conditions,   at one-loop order, the holonomy potential is
\begin{align}
V_{\textrm{eff},  \pm }(P)  & =   
\frac{m}{L \pi} \sum_{n = 1}^{\infty}    (\pm 1)^{n}  \frac{ K_1( n m L)}{n }  |\tr(P^n)|^2      \cr 
&  \xrightarrow[{m \rightarrow 0}]{}  \frac{1}{\pi L^2}\sum_{n=1}^\infty \frac{(\pm 1)^{n}}{n^2}|\tr (P^n)|^2   
\label{hol-pot} 
\end{align}
where $K_1( x)$ is modified Bessel function. 
In the thermal case, the mass-square for Polyakov loop is negative, and   at the one-loop order, the $\Z_N^{0}$ center-symmetry is spontaneously broken. For the thermal case, we show the mass dependence of the potential in Fig.~\ref{Hol-m}. In particular, since only the matter fields contribute to the potential in two dimensions, the potential flattens with increasing mass of the adjoint fermion, $m$. Ultimately, at $m=\infty$, 
pure Yang-Mills theory,  $V(P) \rightarrow 0 $.  In that limit, eigenvalues of the Wilson lines are uniformly distributed.

In thermal case, for generic values of $m$, 
the potential has $N$ degenerate minima, given by 
\begin{align} 
 P_k  = e^{i  \frac{2 \pi k}{N}} {\bf 1},    \qquad \qquad   \theta_1 = \theta_2= \ldots = \theta_N =  \frac{2 \pi k}{N}
 \label{min-broken}
\end{align} 
We can label these perturbative vacua by  $|k \rangle, k=1, \ldots, N$.   For $m>0$,  due to tunnelings between the  perturbative minima,  
 $\Z_N^{0}$  must restore non-perturbatively, as described below.  
  At finite-$N$ and finite $m$,    $\Z_N^{0}$ is broken perturbatively, and restored non-perturbatively due to instanton events.  Indeed, in the simulations, we will exhibits both these degenerate minima as well as the tunneling events between them. 
  
  {\bf Remark:} Since  large-$N$ is a thermodynamic limit, the   $\Z_N^{0}$ center can break, and indeed, it does so  because tunneling rates tends to zero as $e^{-\nc} \rightarrow 0$.  
  Remarkably, in the present theory, similar to charge-$q$ Schwinger model, the massless limit is extremely interesting.  
  Perhaps, up until a few years ago, we would think that the scenario for the $m>0$ case should also hold for for $m=0$, 
  at least for odd-$N$,  where there are no fermionic zero modes  for Dirac operator \cite{Cherman:2019hbq}, and vacuum degeneracy would be lifted because of the instantons. 
  However, the  mixed anomaly (between 1-form symmetry and non-invertible symmetry)  \eqref{algebra} in \AQCD{} and \eqref{ano-s} (between 1-form symmetry and chiral  symmetry)  in Schwinger model  imply  that  
   the transition amplitude between degenerate vacua must be zero.  In \AQCD, this is particularly  strange considering that \eqref{hol-pot}  is just a bosonic potential  with $N$  minima, and there are not robust fermion zero modes for $N$ odd.\footnote{An intuitive way to see  the  non-lifting of degeneracy is as follows.  The \eqref{algebra} can be satisfied by the $N \times N$ clock $C$ and shift $S$ matrices, $CS= \omega SC$,  which act as generators of $\Z_N \times \Z_N$.  Since 
    $\Z_N \times \Z_N$ are exact symmetries, the generators must commute with Hamiltonian, $[C,H] = [S, H]=0$.   But the only thing that commutes with both is identity operator, $H= E {\bf 1}_N$, with exact $N$-fold degeneracy. Hence, degeneracy cannot be lifted despite the existence of instantons.}    But it turns out this is possible. 
   The fact that the transition  amplitude can be zero due to destructive interference related to  bosonic zero modes has been shown in an explicit example \cite{Nguyen:2022lie}. But  the precise way in how the tunneling amplitude must vanish  in the context of \AQCD{} has not yet been shown.
 
For the periodic boundary conditions,  the mass square for Polyakov loop is always positive, and the minimum of the holonomy potential is unique. It is located at the  non-trivial holonomy configuration:
\begin{align} 
    P_{*} &=  e^{ - i a  \frac{\pi}{N}}   {\rm diag} \left( 1, e^{i \frac{2\pi}{N}} , e^{i  \frac{4\pi}{N}  }, \ldots,  e^{i  \frac{2(N-1)\pi}{N} }  \right)  \qquad  a = 0, 1 \qquad  { \rm odd, \; even}  \;  N\; .
    \label{min-sym}
\end{align} 
Therefore, the $\Z_N^{0}$  is unbroken  perturbatively at one-loop order for $m >0$. For $m=0$, despite the fact that   \eqref{min-sym}  is the minimum of the potential, we should expect subtleties related to mixed anomaly. 
 In the simulations, both \eqref{min-broken} and \eqref{min-sym}  comes out rather nicely, see Fig.~\ref{fig:nc6phase}. 

Analytically, one loses control over the one-loop potential when  $g L \gtrsim  O(1)$. It is expected that the eigenvalues of the Polyakov loop will be uniformly distributed in that regime with large fluctuations.   In simulations, we show this by evaluating the expectation value for the modulus of Polyakov loop, and as well as its susceptibility.  In the  thermal $\mathbb R \times S^1 $ set-up, 
there is a cross-over rather than phase transition at finite-$N$, but this becomes a genuine phase transition in the $\nc=\infty $ limit.

\begin{figure}[t]
\begin{center}
\includegraphics[angle=0, width=0.8\textwidth]{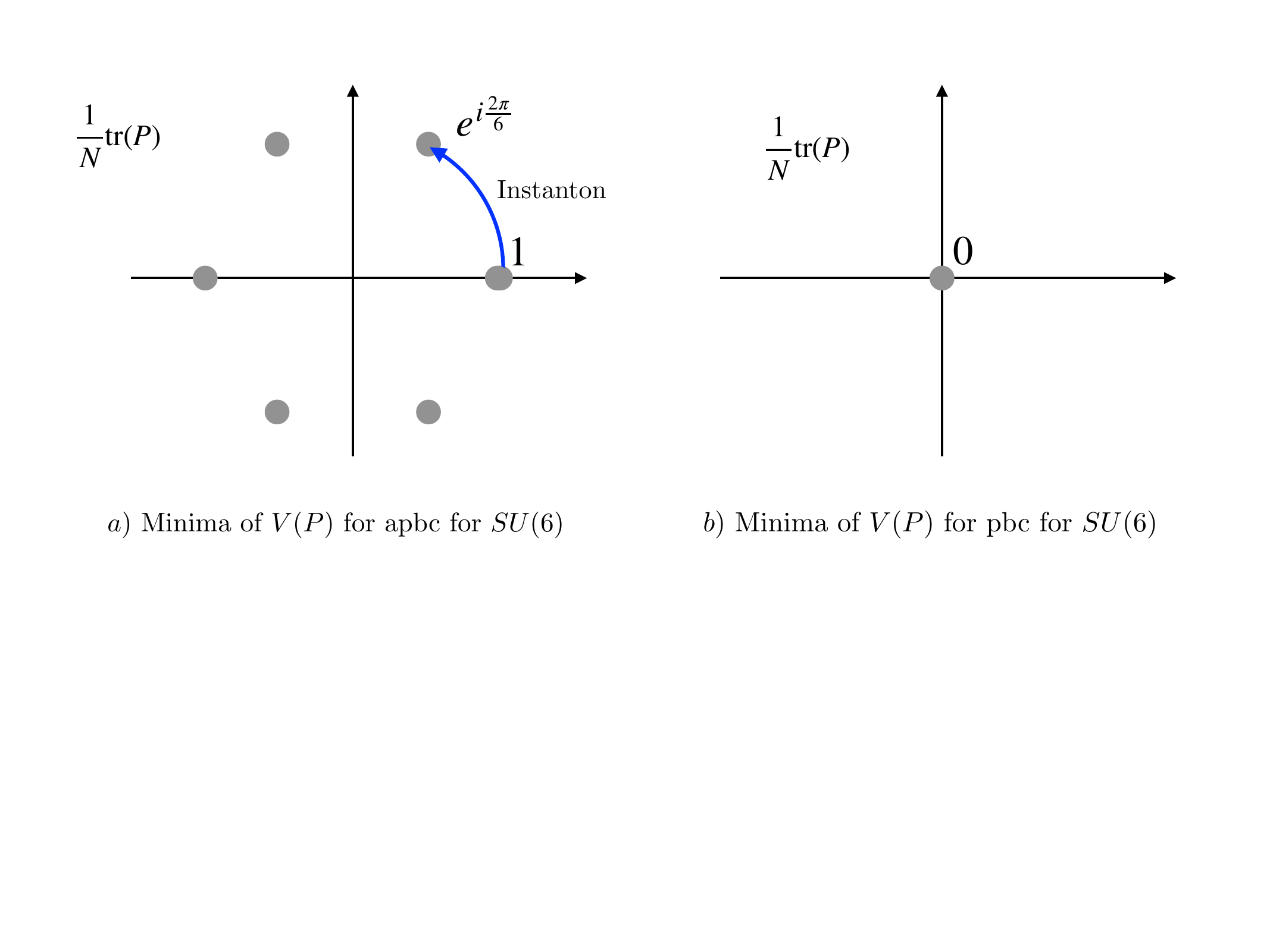}
\vspace{-4cm}
\caption{Mimima of the holonomy potential a) For fermions endowed with anti-periodic boundary conditions, there are N-minima. b) For fermions endowed with periodic boundary conditions, there is unique minimum. }
\label{Wilson}
\end{center}
\end{figure}

 \subsection{Instantons in the thermal case} 
As stated above, in the thermal case, there are $N$-degenerate minima at perturbative level.  To describe the tunnelings between them,   we consider the 1D quantum mechanical action for the Polyakov loop. To understand the nature of these tunneling events, 
we can momentarily ignore  the fermions, which are gapped due to anti-periodic boundary conditions.  The 1D quantum mechanical action is given by  
 \begin{align}
S_{\text{1D}} &= \int \diff t \left[
\frac{1}{2g^2 \beta }\sum_{i=1}^N \dot\theta_i^2 + \frac{1}{\pi L}\sum_{n=1}^\infty \frac{(-1)^n K_1(nLm)}{n^2} \sum_{i,j} \cos n (\theta_i - \theta_j)
 \right] \, .\label{eq:lag1D}
\end{align}
The perturbative minima of this action is given in \eqref{min-broken}, and we would like to determine the tunneling events.  The minimal action configuration interpolating between $|k\rangle$ and  $|k+1\rangle$ can be found by using  
  an abelian ansatz for the Polyakov loop 
 \begin{align} 
 P_{\rm ab.an.} =  {\rm diag} \left(  e^{i  \theta_1} , e^{i  \theta_2 }, \ldots,  e^{i  \theta_N}  \right) =    {\rm diag} \left(  e^{i  \alpha} , e^{i  \alpha }, \ldots,  e^{-i (N-1)  \alpha }  \right) 
 \end{align}
In this parametrization, the action takes the form: 
\begin{align}
S_{\text{1D}} &= \int \diff t \left[\frac{N^2(N-1)}{2(g^2N)L}\dot\alpha^2 
 + \frac{1}{\pi L}\sum_{n=1}^\infty \frac{(-1)^n}{n^2} K_1(nLm)  2(N-1)\cos N\alpha n   
\right] \nonumber \\ 
&  \xrightarrow
    [{m \rightarrow 0}]{}  \frac{N^2(N-1)}{2 \lambda L} \int \diff t \left[
 \dot\alpha^2 +   \omega^2  \min_{k\in\mathbb Z}\left(\alpha +{ \frac{2\pi k}{N} } \right)^2    \right]  \,  .
 \label{eq:lagrangian_anti_periodic}
\end{align}
where $ \omega^2 = \frac{\lambda}{\pi} $.  The potential is harmonic and has cusps at  $\alpha= \pm \frac{\pi}{N}$ etc.  Since the instanton equation is extremely simple,  just motion in the periodic extension of the inverted simple harmonic oscillator, 
\begin{align}
\dot \alpha =  \pm \omega \alpha  \qquad  {\rm for} \;  |\alpha| \leq  \frac{\pi}{N}, 
\end{align}
the full solution can be found by appropriate patching of the solutions  $\exp[ \pm \omega t] $.  \footnote{We do express the solutions in the $m=0$ theory to have a sense of the instantons, and their zero mode structure. At $m >0$, the cusps in the potential  and the cusp in $\dot \alpha(t)$ smoothen.   Moreover, the fermi zero modes will be lifted by soaking them up with the mass operator.  }
The  configurations  interpolating between $\alpha=0$ ($P = {\bf 1}$)  and  $\alpha= \frac{2 \pi}{N}$  
($P = {\bf 1} e^{i \frac{2 \pi }{N}}$)  are given by 
\begin{align}
\alpha(t) =   \frac{2 \pi}{N}  \Theta(t) -  \frac{\pi}{N} {\rm Sign}(t)  \exp[-  \omega |t|]  = 
\begin{cases}\displaystyle
\frac{\pi}{N} \exp[ \omega t]  & t <0,  
\\  \\ 
\displaystyle
 \frac{2 \pi}{N}  -  \frac{\pi}{N} \exp[- \omega t]  & t >0, 
\end{cases} 
\label{eq:inst}
\end{align}  
where $\Theta(t)$ is step function and $ {\rm Sign}(t) $ is sign function. We set the position moduli of the instanton to zero, $t_0=0$.    This is a smooth continuous function extrapolating between the two adjacent minima.  Its derivative is also continuous with a cusp at $t=0$, $\dot \alpha(t) = \frac{\pi}{N} \omega   \exp[-  \omega |t|]  $.    The action of the instanton can be obtained by integrating twice the kinetic term over Euclidean time: 
\begin{align}
S_I =   \frac{N^2(N-1)}{2 \lambda L} \int \diff t \;  2 \;  \dot\alpha^2  = 
  (N-1) \frac{\pi^{3/2}}{ L \lambda^{1/2}}.
\label{eq:instanton_action}
\end{align}
Therefore, the transition amplitude between adjacent vacua (assuming small non-zero mass for fermions) is approximately given by \footnote{
We calculated the instanton action for the holonomy potential at $m=0$. This action remains the same as long as $mL \ll 1$ at leading order. When    $mL \gg  1$, the potential flattens as in  Fig.~\ref{Hol-m}, and  instantons disappear (or ``melt"). The strict $m=0$ limit is more subtle for two reasons. One is that  there is a mod 2 index theorem, which indicates $n_L= n_R=1$ (mod 2) zero mode for $N$=even 
and $n_L= n_R=0$ (mod 2) for   $N$=odd. The action of the instanton is still the same, but even if there is no fermi zero modes (as in $N$=odd), the tunneling must be forbidden because of mixed anomaly between $G_{\rm non-inv}$ and $\Z_N^{[1]}$ according to  \eqref{algebra}. } 
\begin{align}
\langle k+1|e^{-T H} | k\rangle  \sim e^{- (N-1) \frac{\pi^{3/2}}{ L \lambda^{1/2}}}
\label{transition}
\end{align}
Two remarks are in order. The instanton action does not scale as $ \frac{1}{\lambda} = \frac{1}{g^2\nc}$,  rather it is 
$ \frac{1}{\lambda^{1/2}}$. This is due to the fact that this instanton is induced by the balancing of the classical kinetic term with  one-loop potential. i.e.\ it is some sort of quantum instanton. The second point is the action of the instanton scales as $e^{-\nc/L}$, hence these tunneling events become forbidden in the large-$\nc$ limit.    

\begin{figure}[t]
\begin{center}
\includegraphics[angle=0, width=0.8\textwidth]{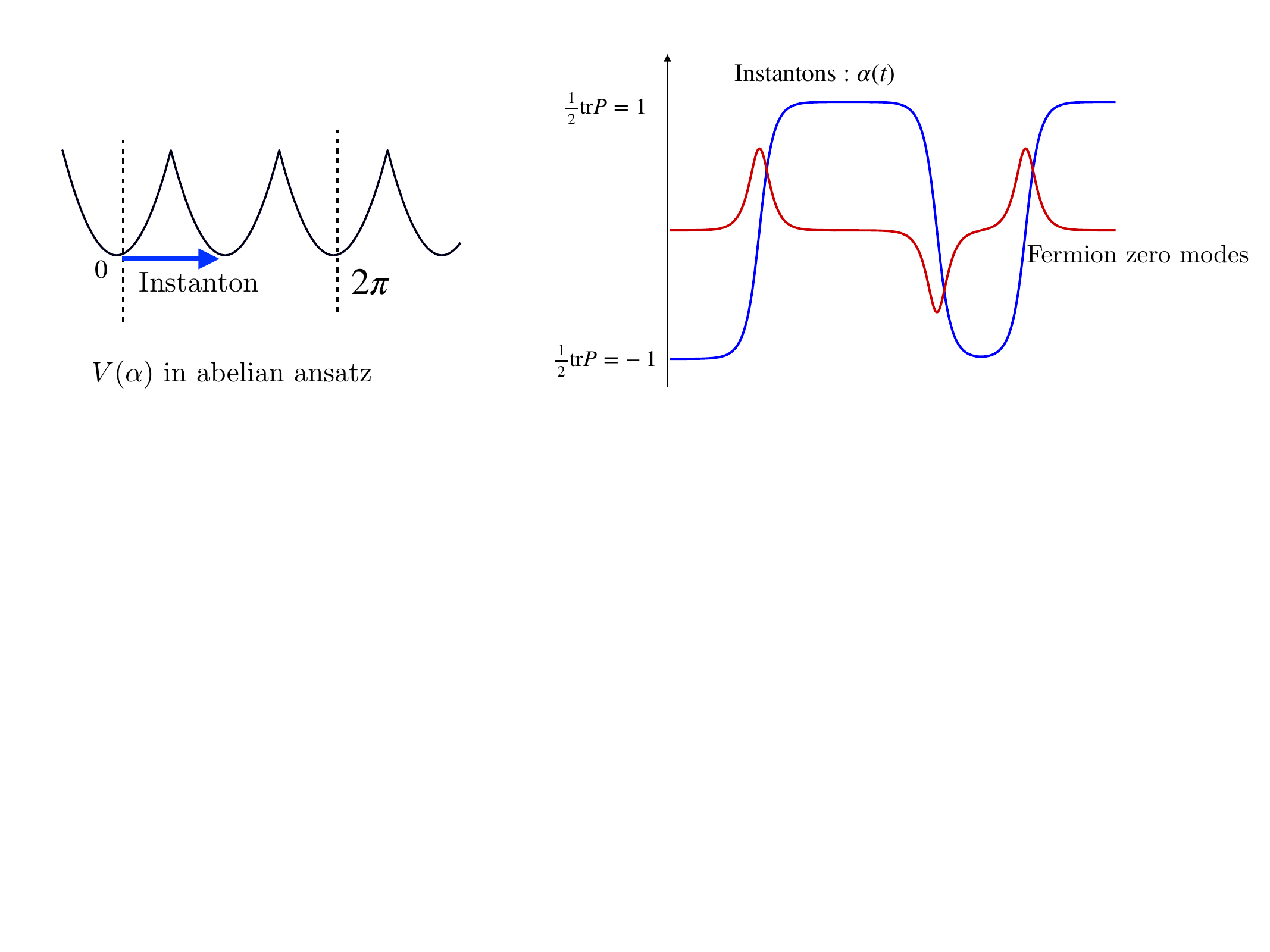}
\vspace{-5.5cm}
\caption{In thermal case, the tunneling between perturbative minima of the holonomy potential 
is associated with one positive and one negative chirality fermion zero mode for even $\nc$.  
For odd $\nc$, these modes are not robust.  In the weak coupling semi-classical domain, the chiral condensate receives its contribution from the fermi zero modes localized on instanton. This is also seen in simulations, see Fig.\ref{fig:condpl}. }
\label{zeromode}
\end{center}
\end{figure}

Non-perturbatively, for finite-$\nc$ and finite $m$,   the degeneracy between the $|k \rangle, k=1, \ldots, N$ states  is lifted because of the  tunnelings. 
 In the Born-Oppenheimer approximation,  the lowest lying $N$ states are 
\begin{align}
|  \Psi_q \rangle = \frac{1}{\sqrt{N}}  \sum_{k=1}^{N} e^{ i \frac{2 pi k q}{N} }  |  k \rangle  
%
\end{align}
where $|  \Psi_0 \rangle$ is the ground state. 
Now, the expectation value of Polyakov loop in the ground state $| \Psi_0 \rangle$   vanishes because of the averaging between these $N$ states; while with the periodic boundary conditions, the bosonic potential has a unique center-symmetric minimum,
 \begin{align}
  \langle \Psi_0  |  \tr P | \Psi_0 \rangle =  \frac{1}{N}  \sum_{q,k}
   \langle q  |  \tr P | k \rangle = \frac{1}{N}  \sum_{k=1}^{N} e^{ i \frac{2 \pi k }{N} } =0\; .
   \label{P-zero}
  \end{align}
In other words, Polyakov loop expectation value  vanishes because of tunnelings in the high-temperature regime. 

In the low temperature, as well as small-$L$ limit  of periodic boundary conditions,  the reason for its vanishing is that eigenvalues of Polyakov loop are uniformly distributed (as opposed to being clumpled), 
 $\frac{1}{N}  \tr P_{*}   =0$.  
  
These two patterns of unbroken center symmetry are physically quite different. In fact, in the large-$\nc$ limit, which is a way to achieve a thermodynamic limit in the context of quantum mechanics, the center-symmetry breaks even non-perturbatively because 
the suppressing of tunneling  as $e^{-\nc} \rightarrow 0$.  For the periodic boundary conditions, since center-symmetry is unbroken, 
large-$N$ volume independence must hold similar to 4d QCD(adj) \cite{Kovtun:2007py}.   This implies that the dynamics of \AQCD{} at $\nc \rightarrow\infty$ on $\R^2$ can  be captured 
by quantum mechanics at small $S^1 \times \R$.  
While for anti-periodic case, volume independence does not hold, and this theory in the large $N$ limit is expected to have a  genuine phase transition at some scale dictated by the strong scale of the theory 
$T_c \sim  \lambda^{1/2} $.

\subsection{Chiral symmetry on $\R \times S^1$} 

\vspace{0.0cm}
\noindent 
{\bf  Even $\nc$, anti-periodic boundary conditions:}
On $\R \times S^1$  with anti-periodic (thermal) boundary conditions,   perturbatively, we have $\nc$ degenerate minima, with instanton configurations interpolating in between. In this case, 
there is a mod 2  index theorem for Dirac operator associated with Majorana fermions, which states that number of left/right handed zero modes modulo 2 is a topological invariant.  
 For $\nc=$even, there exists only one robust pair of zero modes of opposite chirality,   a $\psi_{+ }$ and a $\psi_{-}$, and this can indeed  generate  a chiral condensate via instanton effects \cite{Cherman:2019hbq}.  In the past, it was  thought   that that the  instantons are accompanied with $\nc-1$ positive chirality and $\nc-1$ negative  chirality
 fermion zero modes  \cite{Smilga:1994hc,Gross:1995bp}. As argued therein, except for $\nc=2$, this number is way too large to induce a fermion bilinear condensate in the semi-classical regime. However, as noted above, the mod 2 index theorem resolves this issue  for even $\nc$ \cite{Cherman:2019hbq} (see also \cite{Smilga:2021zrw}.).  
 
Below,   we will describe what happens in detail  for $\nc=2$ and comment for general  even $\nc$. Then, we will discuss odd  $\nc$ case.   For even $\nc$, in particular $\nc=2$,   the mixed anomaly between the chiral symmetry and center-symmetry  survives,
  \begin{align} 
{\mathsf U_{\mathsf s}} (x)  {\mathsf U_{\mathsf \chi}}  (C)   = (-1)   {\mathsf U_{\mathsf \chi}}  (C)   {\mathsf U_{\mathsf s}} (x)   \qquad \text{ for even }\;  \nc\; .
\label{algebra2}
\end{align}
We can view large-Wilson loops  as the generator of the discrete chiral symmetry, and chiral bilinear $\psi_{+} \psi_{-}$ as the generator of the 1-form center-symmetry.  Therefore, in the compactified theory, we can express the algebra in terms of Polyakov loops and chiral operator as: 
  \begin{align} 
\tr \psi_{+} \psi_{-}   \;   P   = (-1)  P  \; \tr \psi_{+} \psi_{-}      \qquad \text{ for even }\;  \nc\; .
\label{algebra3}
\end{align}
For $\nc=2$, the non-invertible symmetry does not play a role, and the story is quite similar to charge-q Schwinger model with $q=2$ \cite{Anber:2018jdf, Anber:2018xek, Armoni:2018bga, Misumi:2019dwq, Honda:2021ovk}. 

We can describe the states in the two dual bases. One is 
the eigenbasis $ |{\ell } \rangle$  of the Polyakov loop  and the other is the eigenbasis  $\widetilde{  |{m } \rangle  } $ of the chiral operator.   The action of the Polyakov loop and chiral operator in these bases are (still focusing on SU(2) in the following): 
\begin{align}
\half \tr P |{\ell } \rangle & = (-1)^\ell   |{\ell } \rangle,  \qquad   \tr \psi_{+} \psi_{-}   |{\ell } \rangle =    |{\ell }+1 \rangle,   \qquad  \ell=0,1 \\
 \tr \psi_{+} \psi_{-}    \widetilde{  |{m } \rangle} &=  (-1)^m  \widetilde{  |{m } \rangle  },  \qquad   \half \tr P  \widetilde{  |{m } \rangle} =    \widetilde{  |{m +1} \rangle  },   \qquad  m=0,1\; .
\end{align}
As a consequence of the algebra \eqref{algebra3}, the $P$ and  $\psi_{+} \psi_{-}$  operators cannot be simultaneously diagonalized. A simple transformation relates the two bases:
\begin{align} 
\widetilde{  |0  \rangle } =  \frac{1}{\sqrt{2}}    ( |0 \rangle+   |1 \rangle), \qquad   \widetilde{  |1  \rangle }=  \frac{1}{\sqrt{2}}    ( |0 \rangle-    |1 \rangle).
\end{align}   

In the $m=0$ case, 
the tunneling amplitude between the eigenstates of the Polyakov loop $  |0 \rangle,  |1 \rangle$  is zero because instantons carry robust  fermion zero modes. Reciprocally, the tunneling amplitudes between the eigenstate of the chiral operator  
$\widetilde{  |0  \rangle }$ and $\widetilde{  |1  \rangle }$ is zero because these states are separated by a (non-dynamical) Wilson line  and belong to distinct universes. On the $\R^2$ limit, the would be domain-wall in this theory carries charge-1 under the $\Z_2$ center-symmetry. But this is not a  dynamical degree of  freedom in the theory, rather an external probe.   This is the reason that 
$\widetilde{  |0  \rangle }$ and $\widetilde{  |1  \rangle }$ are called universes. As a result, we have 
\begin{align} 
  & \langle 1 | e^{-T H } |0 \rangle =0, \qquad     \langle 1 | e^{-T H }  {\mathsf U_{\mathsf s}}   |0 \rangle \neq 0, \\
&\widetilde{    \langle 1 |}  e^{-T H } \widetilde{  |0 \rangle} =0, \qquad   \widetilde{   \langle 1 |} e^{-T H }  {\mathsf U_{\mathsf \chi}} 
\widetilde{    |0 \rangle}  \neq 0\; .
\end{align}

Although instanton in the basis  $| \ell \rangle$  does not lead to tunneling, it causes chiral condensate. As a result, we can express chiral condensate as:
 \begin{align} 
\langle  \tr  \psi_{+} \psi_{-} \rangle = \left\{  \begin{array}{ll} 
 C_1  \frac{1}{ L^2 \lambda^{1/2}}  e^{-  (\nc-1) \frac{\pi^{3/2} }{ L  \lambda^{1/2}}} &\qquad    L  \lambda^{1/2}  \lesssim 1   \\ \\
 C_2 \lambda^{1/2}  &  \qquad  L \lambda^{1/2}  \gg  1  \; .
 \end{array}  \right. 
 \label{chiral}
 \end{align} 
The chiral condensate is non-zero at any temperature, but it is exponentially small in the weak coupling semi-classical regime. The lattice results we obtain are nicely consistent with this structure, much smaller at high-temperature than the low-temperature. This is explained in Sec.~\ref{sec:obs_cond}, see 
Fig.~\ref{sfig:CondTdep}. 

At high-temperature, our lattice results indicate that  the chiral condensate receives dominant contribution from the instanton cores in the semi-classical regime,  in accordance with semi-classical analysis and anomaly structure shown in Fig.~\ref{zeromode}.
See  Fig.~\ref{fig:condpl} for instanton profile on the lattice and its contribution to chiral condensate (further details in Sec.~\ref{sec:obs_cond}).   
    
For even $\nc>2$, we note a few  differences without going into details. First of all, the mode 2 index theorem tells us that $n_{+}=n_{-}=0$ (mod 2)  for 
instantons between $| \ell \rangle \rightarrow  | \ell + 2 \ell'  \rangle $ ($\ell'=1,2,\ldots$), evenly separated tunneling events.  This implies that there is no obtruction for the  transitions between  
$| \ell \rangle \rightarrow  | \ell + 2 \ell'  \rangle $ from the fermi zero modes.
However, the 
mixed anomaly between the non-invertible symmetry and center-symmetry \eqref{algebra} enforces that the tunneling 
between $| \ell \rangle \rightarrow  | \ell + 2 \ell'  \rangle $ must also vanish.  Hence, the ground state  must retains an $\nc$-fold degeneracy.  If one explicitly breaks the $G_{\rm non-inv}$ but not ordinary symmetries, then the ground state will be exactly 2-fold degenerate for even $\nc$, with the value of the condensate as in \eqref{chiral}.

  \vspace{0,3cm}
\noindent 
{\bf  Odd $\nc$, anti-periodic boundary conditions:}
  For $\nc$ odd,   the  mod 2 index theorem indicates that there are no robust fermion zero modes for instantons interpolating between the perturbative vacua $|\ell \rangle, \;  \ell=0,1, \ldots, \nc-1$.   
   The would-be zero modes of the instanton will be lifted by fluctuations. Therefore, one may expect that the chiral condensate must vanish in the thermally compactified theory. We will show that this is indeed the case. However, in the lattice simulations, we observe 
   a finite condensate. Yet, this is not a contradiction as we explained below, because of subtleties related to cluster decomposition. 
   
 We first note one of the remarkable features of the theory. Clearly, the semi-classical potential \eqref{hol-pot} has $\nc$ minima. 
 But because of the absence of fermi zero modes associated with the instantons interpolating between these vacua, one may reasonably think that vacuum degeneracy ought to be lifted. However, the fact that the mixed anomaly   between non-invertible symmetry and 1-form symmetry
 \eqref{algebra} persists in thermal compactification, these  degeneracies  between vacua cannot be lifted. 
 Therefore, quantum mechanically, $\nc$ fold degeneracy must survive. This is indeed quite strange, but true.  The mechanism through which this may happen is not shown in this work, we expect it to be analogous to  the example of the $\mathbb {CP}^{N-1}$ model with winding theta, where degeneracy in a similar situation is not lifted due to subtle effects related to instanton moduli space    \cite{Nguyen:2022lie}.

  Let us work with a specific case, $\nc=3$. 
  In this case, on $\R^2$, there are 
  $2^{\nc-1} = 4$ vacua. 
  These 4 vacua are distributed to three universes with charge $q={0,1,2}$ sectors as  ${2,1,1}$.    
  Below, we start with the states as given in Eq.~(6.28) of Ref.~\cite{Komargodski:2020mxz}.  These states are in the \AQCD{} in which   $(-1)^F : \psi \rightarrow -\psi$ is gauged.  This means 
  one works with a generalized partition function in which one sums over spin structures. 
  Ref.~\cite{Komargodski:2020mxz} refers to the model with gauged $(-1)^F$ as bosonic model and the one it is ungauged as fermionic model.   We work with the latter.  
  By undoing $(-1)^F$ gauging,  
  we can move to standard QCD(adj) in which $(-1)^F$ is a global symmetry. This  amounts to identifications $ {\mathsf v}_1 \equiv {\mathsf v}_2, {\mathsf v}_3 \equiv {\mathsf v}_4$ in their  Eq.~(6.28). 
  The states on $\R^2$ and the expectation value of some operators are therefore given by:  
  \begin{center}
\begin{tabular}{|c | c| c| c|c|} 
 \hline
  & $ {\cal U}_0  $  & $ {\cal U}_0  $   &   $ {\cal U}_1 $ & $ {\cal U}_2  $  \\ 
       $\langle {\mathsf v}_a  | {\cal O}  | {\mathsf  v}_a  \rangle$ & $ {\mathsf v}_1$ & $ {\mathsf v}_3$ &    $ {\mathsf  v}_5$    &   $ 
       {\mathsf v}_6$    \\ 
 \hline 
   ${\cal O}_{\cal A} $ & 1 & 1 & $e^{2 \pi i /3}$ &  $e^{4 \pi i /3}$\\ 
 \hline
    ${\cal O}_{2} $  & -2 & -2 & 1 &  1 \\
 \hline
    ${\cal O}_{\chi}$  &  $- \sqrt{3} $ &  $ \sqrt{3} $ &  0&  0 \\
 \hline
\end{tabular}
\end{center}
Here, ${\cal O}_{\cal A} $ is the one-form symmetry operator, whose value tells us in which universe we are.     ${\cal O}_{2} $ is 4-fermi operator, and     ${\cal O}_{\chi}$  is the chiral operator. Clearly, in  $ {\cal U}_0  $, chiral symmetry is broken.

The degeneracy between the vacuum  states of distinct  universes cannot be lifted because of anomaly \eqref{algebra}, but  multiple degenerate vacua in a given universe can become non-degenerate with thermal compactification. Let us show how this happens. The degeneracy between the two degenerate vacua in $ {\cal U}_0  $ gets lifted once the theory is compactified on a thermal circle on  $\R \times  S^1$.  Following the notation of   \cite{Komargodski:2020mxz}, 
let us  denote the two chirally broken 
  vacua by  $  | {\mathsf  v}_1 \rangle  $  and $  | {\mathsf  v}_3 \rangle $. Since these are related by spontaneous breaking of an ordinary symmetry with no mixed anomaly with other symmetries,  the domain-wall between these two states will be a regular dynamical object. In particular, if we compactify this theory,  these two states are able to mix with each other. Therefore, in the theory 
  on $S^1_{\widetilde L} \times S^1_{L}$ (compactified on space times  Euclidean time), the space-independent configuration interpolating  between the two  chirally broken vacua is an instanton from the quantum mechanical perspective, with an extensive action proportional to  ${\widetilde L}$.  Therefore, in the thermally compactified theory,   true ground state and first excited states in   
  $ {\cal U}_0  $  are: 
 \begin{align} 
 |\pm \rangle=  \frac{1}{\sqrt 2}(   | {\mathsf  v}_1 \rangle  \pm  | {\mathsf  v}_3 \rangle )
 \end{align} 
 The energy splitting between  $|- \rangle$ and  $|+ \rangle$  is proportional to $e^{- \widetilde L g}$ due to tunneling between these two vacua.   
 In both of these  $|\pm \rangle$ states,  the expectation value of the chiral condensate vanishes, 
\begin{align} 
 \langle +  | \psi_{+} \psi_{-}  |+  \rangle = \textstyle{ \frac{1}{2} } \left( \langle  {\mathsf  v}_1   | \psi_{+} \psi_{-}   |{\mathsf  v}_1   \rangle  + 
\langle {\mathsf  v}_3  | \psi_{+} \psi_{-}   |{\mathsf  v}_3  \rangle \right) =  \textstyle{ \frac{1}{2} } ( \sqrt{3}   -  \sqrt{3}  ) = 0 \; ,
\label{vanish-ch}
\end{align} 
because of the tunnelings. However, note that the tunneling amplitude 
is vanishing exponentially with volume of the space $e^{-{\widetilde L} g}$, consistent with the fact that at ${\widetilde L}  \rightarrow \infty 
$  limit, these two vacua becomes two superselection sectors in ${\cal U}_0$.

In our simulations, we observe a chiral condensate for SU$(3)$ 
at all temperatures. 
It is extremely plausible that, at least in the large $\widetilde L$ limit, our simulation does not tunnel between the two chirally broken vacua as this rate is suppressed by  $e^{-\widetilde L g}$, and also   because we have a bare mass preferring  one vacuum over the other (which is probably even a stronger effect 
as we take $\widetilde L \rightarrow \infty $ limit before we take the chiral 
limit. 
Furthermore, it is not the states  $|\pm \rangle$ that satisfy cluster decomposition on $\R^2$ limit, rather it is the states   
$ |{\mathsf  v}_1   \rangle ,  |{\mathsf  v}_3   \rangle $. Therefore, even though formally, the chiral condensate  must vanish because of  tunnelings as in \eqref{vanish-ch}, if we turn on a bare mass (which we do), we are essentially forcing the theory to one of the states, say, $ |{\mathsf  v}_1   \rangle $ which satisfies cluster decomposition and for which the condensate is non-zero. 
With the above considerations, we end up with three vacua, one in each universe in the large-$\widetilde L$, large-$L$ regime by the compactification of the theory. 

In large-$\widetilde L$, but small-$L$, we  found three perturbative vacua \eqref{min-broken} as well, by performing an analysis for the gauge holonomy potential,    labelled by the phases of the trace of  Polyakov loop.   These two sets of vacua are naturally discrete Fourier transform of one another, as a consequence of mixed anomaly, \eqref{algebra}.  
 \begin{align} 
| \ell \rangle  = \frac{1}{\sqrt {3}} \big( |+ \rangle+   e^{ i \frac{2 \pi}{3} \ell}  |{\mathsf  v}_5     \rangle  +   e^{ i \frac{4 \pi}{3} \ell}  |{\mathsf  v}_6     \rangle   \big) 
\end{align} 
At large-$L$, we can understand the absence of tunneling by the fact that the domain-walls of the theory on $\R^2$ are non-dynamical and carry non-trivial charge under $\Z_3^{[1]}$ one-form  symmetry. Since there are no such charged objects in the theory, we cannot realize these tunneling events dynamically.
 On the dual  eigenbasis $| \ell \rangle$  of the Polyakov loop,   
the tunnelings  must again vanish. However, since the mod 2 index theorem tells us that there are no fermionic zero modes,  we expect  this to  occur because of  integration over the instanton moduli space. 
Certainly, a more in-depth  study is needed concerning these general  issues  for general  $\nc$.

\section{Observables on the lattice}
\label{sec:obs}
The relevant observables to understand the bevaviour of the theory as a function of compactification radius  or inverse temperature  on $\R \times S^1$ are  
 the Polyakov loop $P_L$, its modulus, susceptibilities, as well as   
the chiral condensate $\psb\psi$.
The volume averaged Polyakov loop is defined as
\begin{equation}
 P_L = \frac{1}{\nc N_x} \sum_{x} \textrm{Tr}\left\{\prod_{t=1}^{N_t} U_1(x,t) \right\}\, ,
\end{equation}
where $U_1$ are the lattice links in $t$ direction (defining the temperature $T/a=1/N_t$). However, as stated in the theory discussion, 
the expectation value of the Polyakov loop vanishes both at high-$T$ (due to tunneling), as well as low-$T$ due to uniform eigenvalue distribution.  Therefore, it is also useful to inspect the modulus of the Polyakov loop, as well as its susceptibility which are clearer indicators of the temperature dependence.  
We also find it useful to examine the scatter plots of trace of Polyakov loops above and below the cross-over temperatures,  as well as   profiles of the Polyakov loops  as a function of the Monte-Carlo time.  

The chiral condensate represented here corresponds always to the volume averaged value,
\begin{align}
-<\psb\psi>=<\frac{1}{N_s N_t}\Tr D_L^{-1}>
\end{align}
with the lattice Dirac operator $D_L$. With Wilson fermions the condensate is subject to additive and multiplicative renormalization. However, since the renormalization is temperature independent, the temperature dependence can still be directly observed even from the Wilson-Dirac operator. To determine the  zero temperature value  of the condensate, we use overlap fermions by reweighting, see  Sec.~\ref{sec:zeroT_cond}. These fermions have much better chiral properties. 

At finite temperature,  the chiral condensate is expected to be non-zero 
at any temperature, at least for even-$N$. This is also true for odd-$N$ 
for a state which would satisfy cluster decomposition at $\R^2$ limit, but not for symmetric state. See the discussion in the previous section. 
As shown in \eqref{chiral}, the chiral condensate is of order of the strong scale of the theory  at low temperatures.  At high temperatures, it is  expected to be exponentially suppressed,   and  dominated by instanton effects \cite{Smilga:1994hc,Cherman:2019hbq}.  

Susceptibilities are also useful indicators of scales  at which the characteristic of an observable changes, and how it changes.  
For observable $O$, it can be defined as:
\begin{equation}\label{eq:suscept}
 \chi_O = \tilde{V} (\langle O^2 \rangle - \langle O \rangle^2)\,,
\end{equation}
where $\tilde{V}$ is $N_x$ for the Polyakov loop or $N_sN_t$ for the chiral condensate. Note that the Polyakov loop susceptibility $\chi_{P_L}$ refers, unless stated otherwise, to the susceptibility of the modulus of the Polyakov loop. 

Some operators in this theory, which are important for later discussions of bound states and parameter tuning are
\begin{align}
\label{eq:op_meson}
    P_\psi=\bar{\psi}\gamma_\ast\psi\;,\quad 
    S_\psi=\bar{\psi}\psi\; .    
\end{align}
As explained Sec.~\ref{sec:parameter_tuning}, we can introduce an effective mass $m_\pi$ based in the operator $P_\psi$ removing the additive mass renormalization of Wilson fermions.

We investigate the string tensions by  evaluating the potential between a pair of test quarks as a function of fermion mass. We apply several methods to obtain more reliable estimates as explained in Sec.~\ref{sec:string_tension}. Finally, we also consider the lightest bound states of the theory. First preliminary results are discussed in Sec.~\ref{sec:light_states}.

\subsection{Polyakov loop related observables}
\label{sec:obs_pl}
\begin{figure}
	 \centering
\subfloat[thermal bc\label{fig:nc6phasea}]{\includegraphics[width=0.48\textwidth]{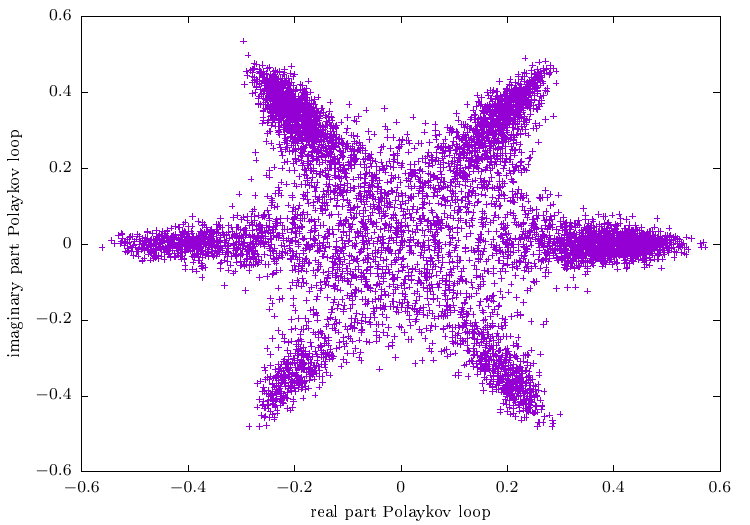}}
\subfloat[periodic bc\label{fig:nc6phaseb}]{\includegraphics[width=0.48\textwidth]{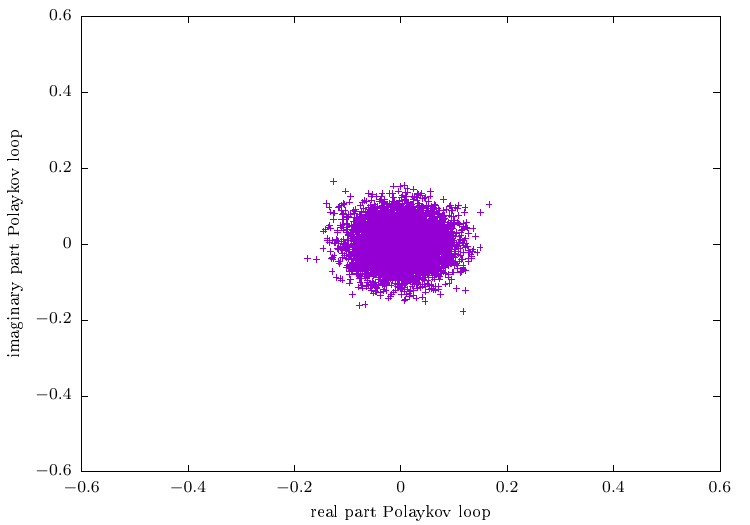}}
	\caption{SU(6) \AQCD\ on a $4\times 16$ lattice at $\kappa=0.265$, $\beta=160$ ($am_\pi=0.2491(15)$). The distribution of real and imaginary part of the volume averaged Polyakov loop for each configuration is represented in a scatter plot. Polyakov loop for two different fermion boundary conditions at small-circle (high-$T$) are shown: thermal bc (antiperiodic for fermions) and periodic bc (periodic for all fields).   Also note that at large-circle, the scatter plot for both boundary conditions tends towards pattern (b).
 At small-circle, again the same tendency occurs with increasing the mass of the fermion, as the potential flattens as in Fig.~\ref{Hol-m}.
 }
	\label{fig:nc6phase}
\end{figure}

In Fig.~\ref{fig:nc6phase}(a), we show the scatterplot of the  Polyakov loop of the generated ensemble for $\Su{6}$ gauge theory both for anti-periodic and periodic boundary conditions at  small $S^1 \times \R$ (corresponding to high-temperature in thermal compactification.)  Indeed, the plot shows that real and imaginary part of the Polyakov loop are concentrated in the neighborhood  of six distinct phases, along rays in directions given by $e^{i  \frac{2\pi k}{6} }$ ($k=0,1,\ldots 5$). These are 
the perturbatively broken vacua found in \eqref{min-broken}. The simulation scans all possibilities of these phases, which correspond to the scanning of all minima of the potential via  tunneling events in the analytic formulation. Note that in this case,  despite the fact that center symmetry is broken perturbatively, 
it is unbroken non-perturbatively due to instanton events \eqref{eq:inst}. 
For subtleties related to $m=0$ limit, see the remark in Sec.~\ref{sec:apbc}.

When the fermions are endowed with periodic boundary condition, in the scatterplot Fig.~\ref{fig:nc6phase}(b), we see a distribution around 
$P_L =0$. This implies that $\Z_N$ center-symmetry must remain unbroken at small $S^1 \times \R$, consistent with the theoretical result \eqref{min-sym} at least for fermion mass $m> 0$.   We also note that in the finite-temperature formulation, the scatter plot of the Polyakov loop at low temperatures is essentially identical to Fig.~\ref{fig:nc6phaseb}. This provides already an illustrative picture of the transition, but a characterization of its nature requires more detailed analysis of the scaling.

\begin{figure}
	\centering
 \subfloat[modulus of $P_L$\label{fig:pltransitiona}]{\includegraphics[width=0.48\textwidth]{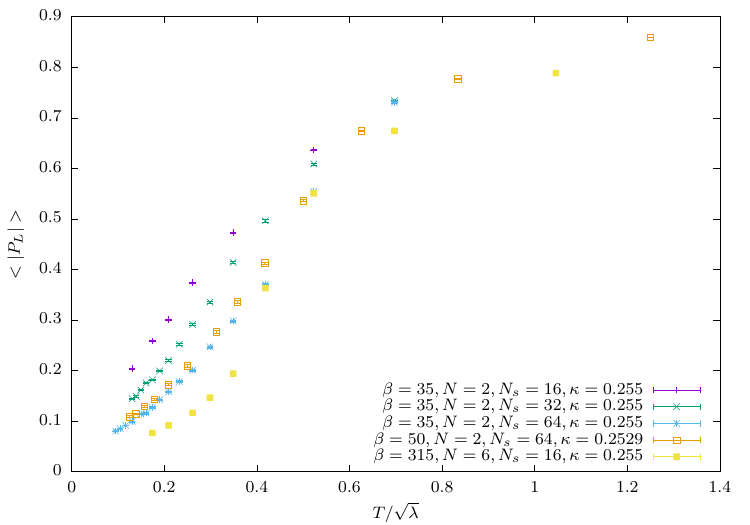}}
  \subfloat[$P_L$ susceptibility\label{fig:pltransitionb}]{\includegraphics[width=0.48\textwidth]{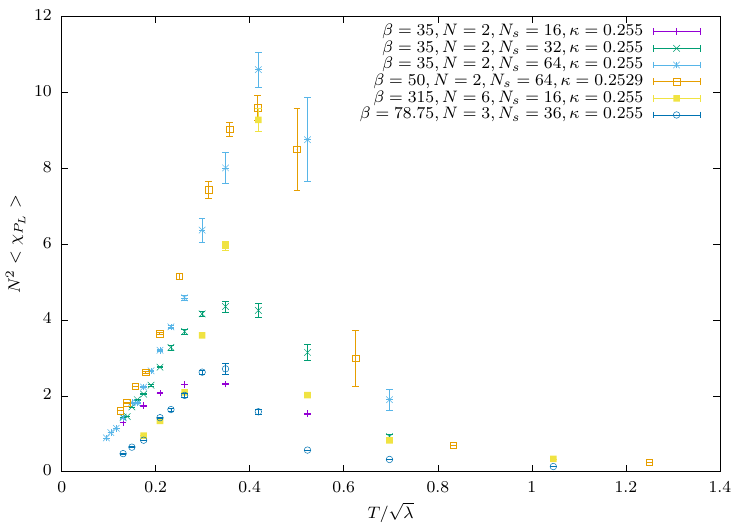}}
	\caption{The average modulus and susceptibility of the volume averaged polyakov loop as a function on an $N_t\times N_s$ lattice with thermal boundary conditions for \Sun{} \AQCD.}
	\label{fig:pltransition}
\end{figure}
As explained around  \eqref{P-zero}, the expectation value of the Polyakov loop is zero even at high temperature, 
because of the tunneling between the perturbative vacua $\tr P = e^{ i \frac{2 \pi k}{N}}$.  At low temperatures, we expect that  the eigenvalues are uniformly distributed, and  there, 
it is  the reason for the vanishing of the Polyakov loop. 
 At high temperatures, eigenvalues clump, but then, they can tunnel collectively among vacua. In order to see this effect, it is more useful to inspect modulus of the Polyakov loop.  
The value of modulus of the Polyakov loop shows a dramatic change  from small to large values  with  the temperature as shown in Fig.~\ref{fig:pltransition}. A transition point can be identified by the peak of the susceptibility, see Fig.~\ref{fig:pltransitionb}). This behaviour is consistent for different $\nc$ at fixed temperatures in units of the 't Hooft coupling $\frac{T}{\sqrt{\lambda}}=\frac{1}{N_t}\sqrt{\frac{\beta}{2\nc^2}}$.  
The modulus of the Polyakov loop is small at low-$T$ and it is approaching to one at high-$T$, with an inflection point around $T \sim 0.2, 0.3 \sqrt \lambda$, where susceptibility reaches to its maximum. 
\begin{figure}
	\centering
 \subfloat[modulus of $P_L$\label{fig:pltransitionNca}]{\includegraphics[width=0.48\textwidth]{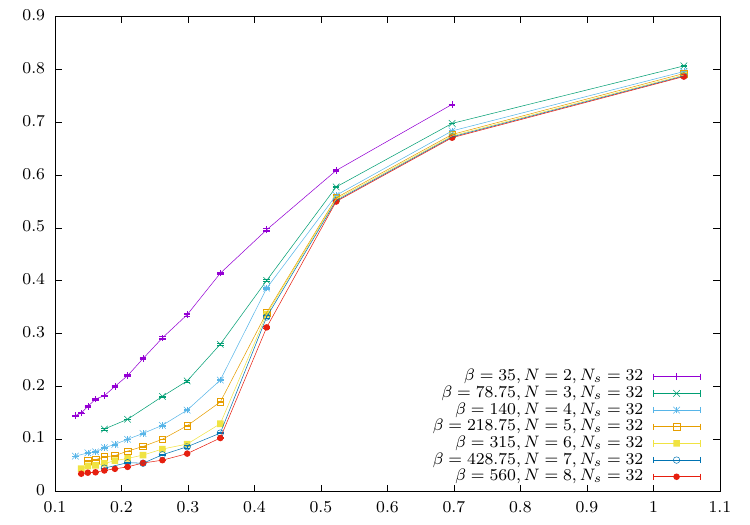}}
  \subfloat[$P_L$ susceptibility\label{fig:pltransitionNCb}]{\includegraphics[width=0.48\textwidth]{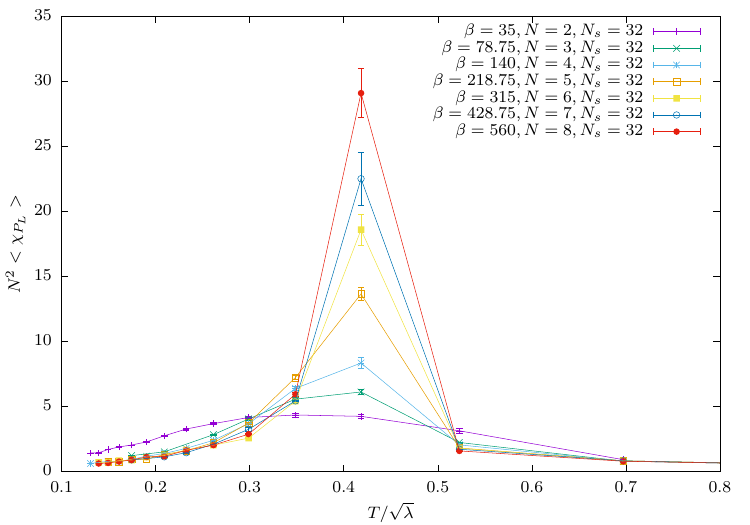}}
	\caption{The scaling of the results presented in Fig.~\ref{fig:pltransition} with $N$.}
	\label{fig:pltransitionNc}
\end{figure}
\begin{figure}
	\centering
 \subfloat[modulus of $P_L$]{\includegraphics[width=0.48\textwidth]{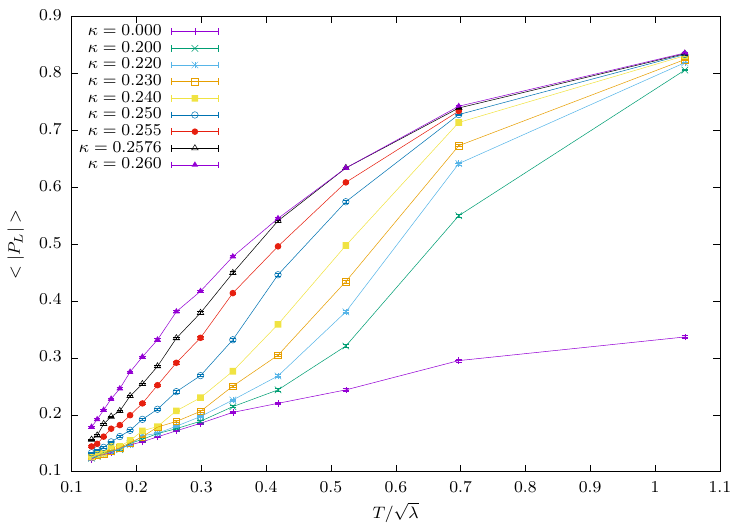}}
  \subfloat[$P_L$ susceptibility]{\includegraphics[width=0.48\textwidth]{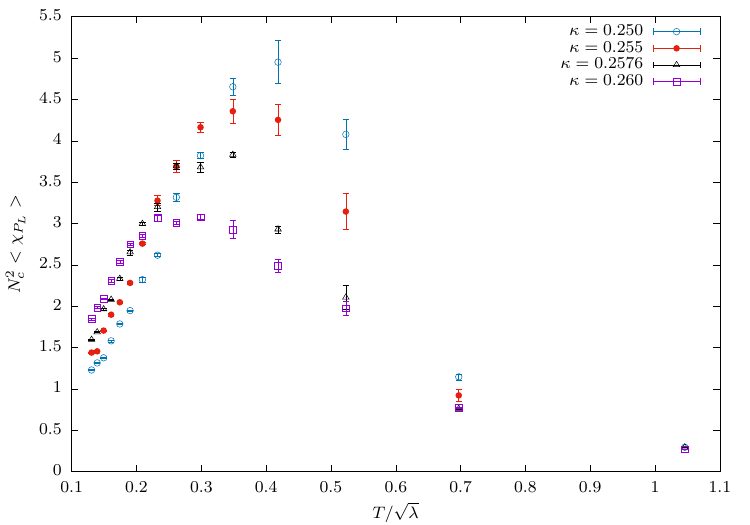}}
	\caption{The mass dependence of the transition of the Polyakov line in \Su{2} \AQCD with $N_s=32$ and $\beta=35$. In (b) the Polyakov susceptibility for a subset of the masses is shown.}
	\label{fig:pltransitionMassDep}
\end{figure}
\begin{figure}
	\centering
    \includegraphics[width=0.48\textwidth]{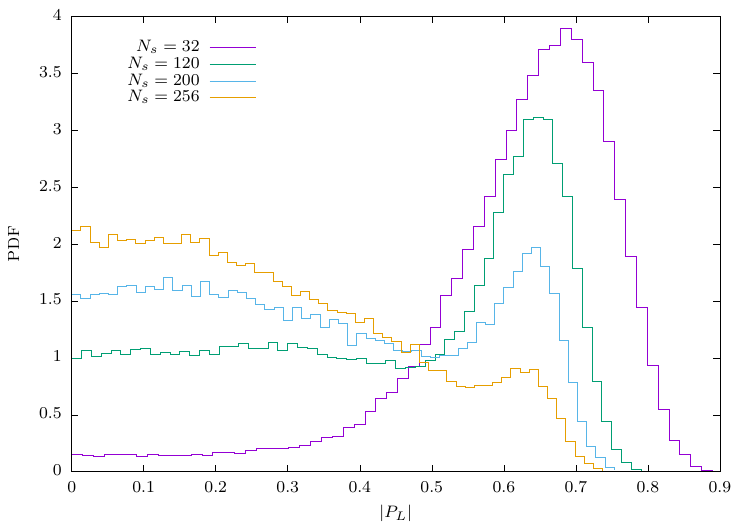}
	\caption{Histogram of the Polyakov loop in SU(2) \AQCD at $\beta=35$ and $N_t=4$ for different $N_s$.}
	\label{fig:pltransitionvdep}
\end{figure}

Although the plot of  susceptibility  has a peak,  and the one of 
the modulus of the Polyakov loop  has an inflection point, there is no genuine phase transition here for finite-$N$ in the large $N_x$ limit.
The decreasing susceptibility at the highest temperatures and the narrowing of the distribution around the center broken vacua observed at a fixed volume does not indicate a transition in this case. The tunneling probability \eqref{transition} and hence the density of tunneling events scales with $e^{-N/L}$ and is hence reduced at large $T$ but independent of $N_x$. 
The effect is rather due to the fact that the average distance of tunneling events increases with the temperature. When it becomes comparable to $N_x$, the lattice size is not large enough to accommodate enough tunneling events between the vacua. Going to the thermodynamic limit by increasing $N_x$, the number of tunnelling events eventually increases, however, linearly since the density is constant. The volume dependence is hence opposite to the usual expectation at a phase transition where tunneling is exponentially suppressed with the volume. In the present case, the tunneling increases at larger volumes and contribution of a deconfined signal in volume averaged observables reduces as shown in Fig.~\ref{fig:pltransitionvdep}. The peak of susceptibility  and inflection point can rather be seen as finite-$N$ remnants of the large-$N$ phase transition.

In this theory, a thermodynamic limit can be achieved at large-$N$, which introduces infinitely many degrees of freedom.   Since  the tunneling at high-$T$ gets suppressed  by $ {\rm exp} [ {-(N-1) \frac{ \pi^{3/2}T}{  \lambda^{1/2}}} ]$ according to \eqref{transition}, the $\Z_N$ center-symmetry breaks spontaneously, as opposed to the finite-$N$ case where tunneling restors it non-perturbatively, and hence, \eqref{min-broken} are realized as vacua non-perturbatively. Therefore, one obtains a  genuine phase transition at large-$N$ limit. 
 \begin{align}
  {\rm lim}_{ N \rightarrow \infty} \langle P_L \rangle = \left\{ \begin{array}{ll}
 0 & \qquad   T< T_c \cr 
 e^{2 \pi i k/N} &  \qquad T >T_c
 \end{array}  \right.
 \end{align}
 Consequently, as shown in Fig.~\ref{fig:pltransitionNc}, the expected scaling of a phase transition is observed in this limit.

In Fig.~\ref{fig:pltransitionMassDep}, we show the mass dependence of the modulus of the Polyakov line  over a temperature range.  The modulus  increases  consistently at lower fermion masses. 
We can understand this from theoretical considerations. In the holonomy potential,  increasing the fermion mass,  suppresses  
the potential exponentially. In \eqref{hol-pot}, the coefficient of the $|\tr P^n|^2$ is Bessel function of the first kind, $K_1 (n L m)$, and for large values of $m$, it vanishes exponentially, i.e., 
\begin{align}
K_1(mLn) \rightarrow \sqrt \frac{\pi}{2 mLn} e^{-nLm}, \qquad {\rm hence} \;  \lim_{ m \rightarrow \infty}
V(P)=0\; .
\label{eq:massscale}
\end{align}
Since the potential is flattened at larger masses, the potential, even at high-$T$, can no longer pin the holonomy $P$ at the roots of unity $P= e^{i \frac{2 \pi k }{N}} {\bf 1}$. This is sensible because the theory reduces to pure Yang-Mills in the limit, which is almost topological.
Hence, even at high-$T$ regime, the eigenvalues start to become uniformly distributed, and  
$\tr P$  tends towards zero.  This is indeed nicely observed in Fig.~\ref{fig:pltransitionMassDep}. Similarly, with increasing mass, the gradually flattened effective potential is reflected by a larger maximum susceptibility and the increased density of tunneling events shifts the peak towards higher temperatures. 

\subsection{Instantons  and chiral condensate at high-temperature}
\label{sec:obs_cond}
\begin{figure}
	\centering
 \subfloat[MC history\label{sfig:MChistCond}]{\includegraphics[width=0.48\textwidth]{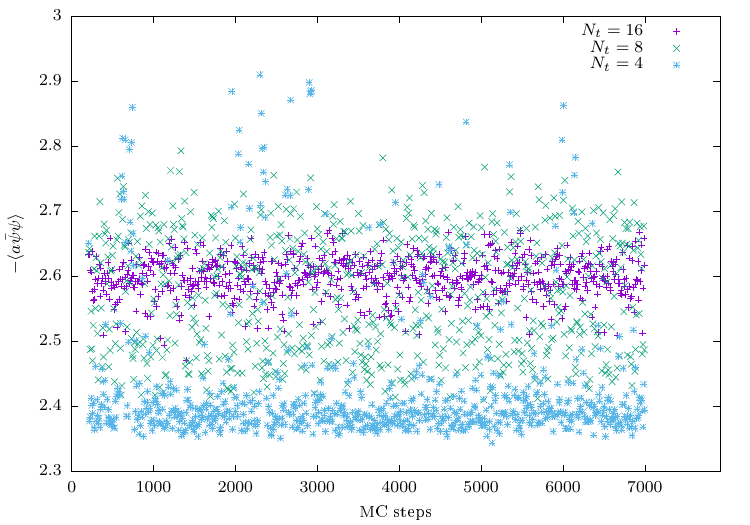}}
  \subfloat[temperature dependence\label{sfig:CondTdep}]{\includegraphics[width=0.48\textwidth]{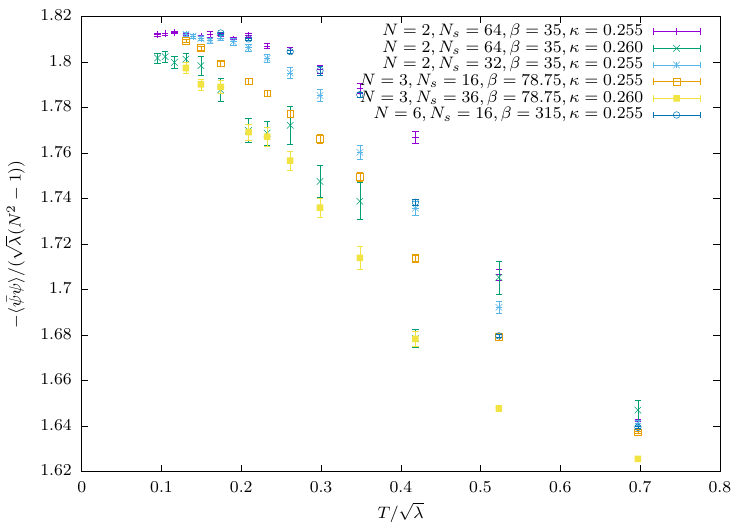}}
	\caption{Temperature dependence of bare Wilson fermion condensate. (a) SU(2) \AQCD on an $N_t\times 32$ lattice, $\kappa=0.255$, $\beta=35$. The fermion condensate distribution of the ensembles at different temperatures. (b) The temperature dependence of the average fermion condensate for different masses and $\nc$ at fixed $\lambda$.}
	\label{}
\end{figure}

In Fig.~\ref{sfig:MChistCond}, 
the Monte-Carlo histories of the chiral condensate are shown as a function of temperature.  At lower temperature (high-$N_t$), the value of the condensate is higher, and it starts to fall-off at higher temperature (low-$N_t$).  The temperature dependence of the condensate is shown in Fig.~\ref{sfig:CondTdep}. 
 Although we cannot calculate the chiral condensate analytically in \AQCD, in strong coupling regime, it is worthwhile to point out that Fig.~\ref{sfig:CondTdep} matches very closely to our analytical guess \eqref{chiral}.  It is interesting to note that \eqref{chiral} 
 carries main features of the chiral condensate in the  Schwinger model as a function of temperature, where in the latter case, it is analytically calculable at any $T$ \cite{Sachs:1991en,Misumi:2019dwq}. 
 In our simulation, note  that due to the mentioned large contribution of a small number of configurations, the error is largely underestimated for higher temperatures and smaller fermion masses.
 
\begin{figure}
	\centering
 \subfloat[MC history]{\includegraphics[width=0.48\textwidth]{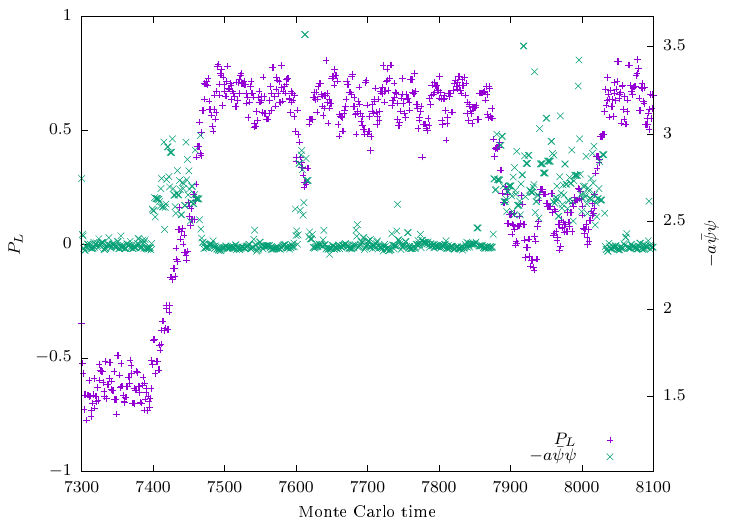}}
  \subfloat[$P_L$ on sample configurations]{\includegraphics[width=0.48\textwidth]{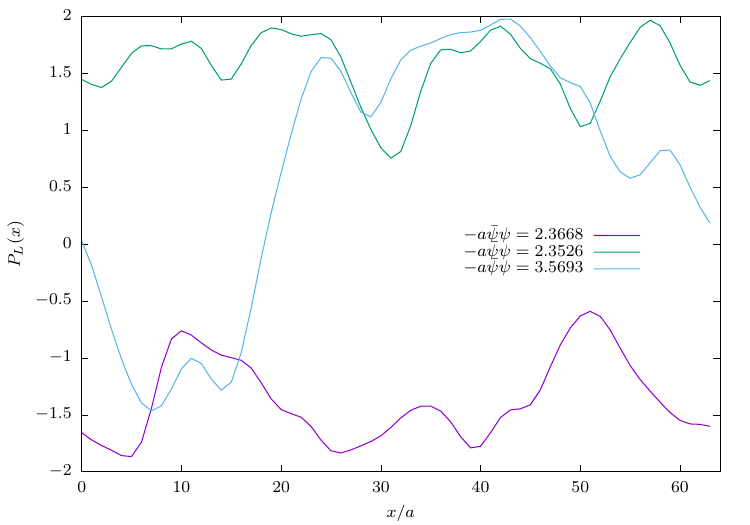}}
	\caption{Correlation between Polyakov loop and condensate in \AQCD{} ($4\times64$ lattice, $\beta=35$, $\kappa=0.260$). (a) The Monte-Carlo history of the Polyakov loop on each configurations is compared to the condensate. (b) The distribution of the Polyakov loop before averaging over $L_0$ for sample configurations. Gradient flow with Wilson action and flow time $\tau=2$ has been applied in (b).}
	\label{fig:condpl}
\end{figure}

\begin{figure}
	\centering
    \includegraphics[width=0.48\textwidth]{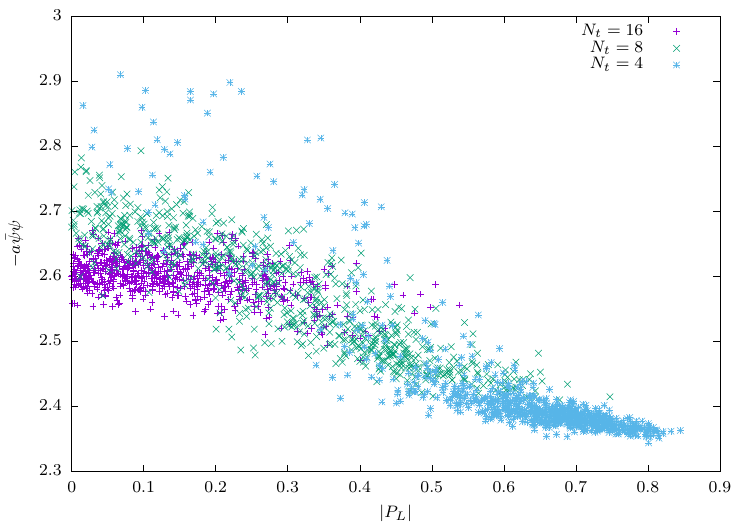}
	\caption{For each configuration the value of the condensate and the Polyakov loop in \AQCD are correlated ($N_t\times 32$, $\beta=35$,$\kappa=0.255$). }
	\label{fig:condplcorrelation}
\end{figure}

Fig~\ref{fig:condpl} is the counterpart of the instanton analysis at high-$T$ in continuum, shown in  Fig.~\ref{zeromode}. 
At high-$T$, 
 there are certain configurations which contribute significantly  larger values  to condensate.  As the semiclassical analysis in \cite{Smilga:1994hc, Cherman:2019hbq} suggests, the larger values of the condensate are related to configurations where the Polyakov loop, taken as a function of the spacial coordinate $x$ without volume averaging, tunnels between the two adjacent minima
of the gauge holonomy potential,  see Fig.~\ref{zeromode}.   In continuum,   a mod 2 index theorem for the Dirac operator of massless Majorana fermions states that for $\nc$ even, there is a robust positive chirality and negative chirality fermion zero mode associated with these tunneling events \cite{Cherman:2019hbq}.  The density (or number) of these configurations is proportional to instanton fugacity $ ~e^{-S_I} \sim e^{ - (N-1) \pi^{3/2} T / \lambda^{1/2} }$
 and reduces with  increasing temperature.
 These observations are shown in Fig.~\ref{fig:condpl} which matches nicely with the semi-classical expectations, shown in  Fig.~\ref{zeromode}. We evaluated the temperature dependence of the condensate for $N=2,3,6$, and do not observe a qualitative difference  between $N$-odd and $N$-even.  
 
 Fig.~\ref{fig:condplcorrelation} shows  the correlation between the chiral condensate and the Polyakov loop. In particular, 
  the chiral condensate is much larger at low-$T$, where the Polyakov loop is disordered and has large-fluctuations but its volume average is small. The condensate is much smaller at high-$T$, 
  where the Polyakov loop is pinned at the minima of the holonomy potential and the condensate receives a dominant contribution from a dilute gas of instantons. These instanton contributions are indicated by the scattered points with small volume average Polyakov loop and large condensate.

\subsection{Zero temperature fermion condensate}

The fermion condensate with Wilson fermion is subject to additive and multiplicative renormalization. The additive renormalization, which makes the interpretation of the zero temperature condensate difficult, is absent with overlap fermions. In our current study, we have not generated ensembles with an overlap operator, but instead we have used reweighting techniques to get the zero temperature condensate. The method based on the complete eigenvalue spectrum of the overlap operator has already been applied in two dimensions for the 't Hooft and Schwinger model. Details are explained in \cite{Berruto:2002gn}. In this reference the reweighting has been done from pure Yang-Mills theory to the ensemble with overlap fermions. In our case, we have applied the same techniques for a reweighting from Wilson to overlap operator. For that purpose, we have calculated the eigenvalues of the hermitian Dirac-Wilson operator and the overlap operator to obtain the ratio of the Wilson and overlap operator determinants as well as the fermion condensate with the overlap operator.

\label{sec:zeroT_cond}
\begin{figure}
	\centering
 \subfloat[different $\nc$ and $\kappa$, fixed $\lambda$]{\includegraphics[width=0.48\textwidth]{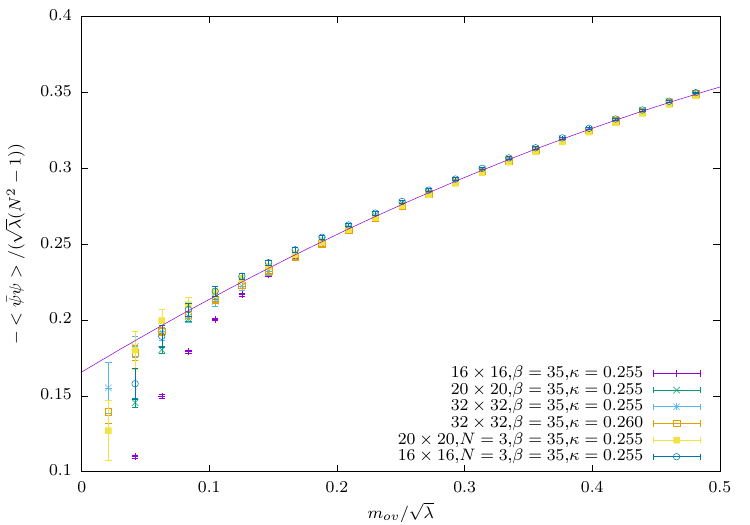}}
  \subfloat[$\beta$ dependence]{\includegraphics[width=0.48\textwidth]{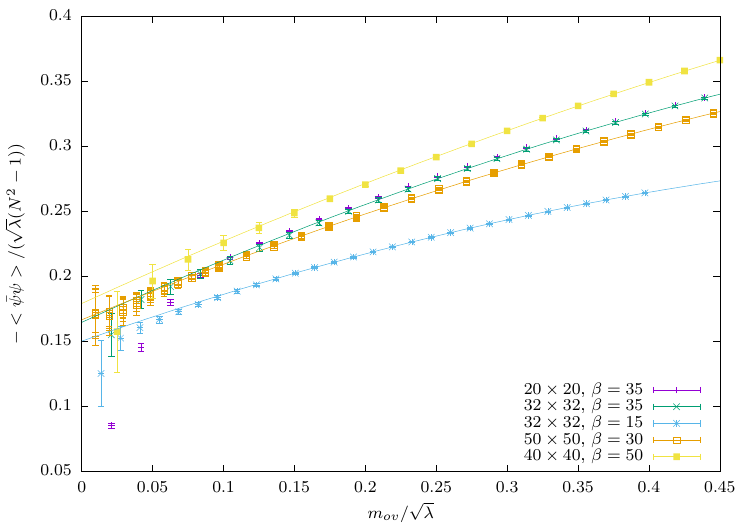}}
	\caption{Zero temperature fermion condensate in \AQCD. The condensate is obtained from a reweighting of Wilson to overlap fermions. The mass $m_{ov}$ corresponds here to the mass in the overlap operator and $\kappa$ to the value of the mass parameter in underlying Wilson ensemble.
 A fit up to quadratic order has been used for the chiral extrapolation.
 (a): At fixed $\lambda$, different $\nc$, $\kappa$, and volumes lead to consistent results. (b): Different $\beta$ lead to approximately compatible values in the chiral limit.}
	\label{fig:fermioncond}
\end{figure}
A meaningful value for the fermion condensate in the chiral (zero mass) limit can only be obtained if the infinite volume limit is taken before the chiral limit. Indeed, the condensate turns to zero if we take the limit of zero fermion mass at a fixed volume. A suitable method is to extrapolate the condensate to zero fermion mass from a region, in which the results for different volumes are consistent with each other. 

The results are presented in Fig.~\ref{fig:fermioncond}. They indicate a finite value of the condensate in the chiral infinite volume limit. The results are independent of the mass parameter ($\kappa$) for the underlying Wilson ensemble, but we observe smaller errors at larger $\kappa$.
With an appropriate scaling, the results of different $\nc$ are consistent with each other at a fixed 't Hooft coupling. The chiral limit seems to be quite independent of $\beta$ given the uncertainties of the extrapolation. However, the scaling with the mass parameter $m$ is quite different for the different $\beta$ and a multiplicative renormalization of the mass parameter appears to be required. It is interesting to note that the fit result in the chiral limit ($-< \bar{\psi}\psi>/(\sqrt{\lambda}(\nc^2-1))=0.1657(33)$)
is close to the Schwinger model ($-< \bar{\psi}\psi>/g=0.15993\ldots$) up to a rescaling of $\sqrt{\nc}(\nc^2-1)$. Note that there is about a factor of two difference with respect the results obtained with a lattice Hamiltonean approach \cite{Dempsey:2023fvm}.

\subsection{String tension}
\label{sec:string_tension}
\begin{figure}
	\centering
\subfloat[$\kappa=0.240$]{\includegraphics[width=0.48\textwidth]{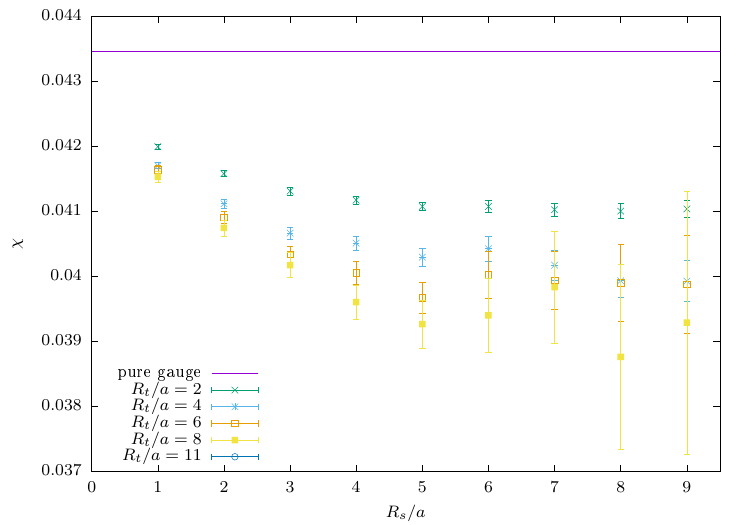}}
\subfloat[$\kappa=0.260$]{\includegraphics[width=0.48\textwidth]{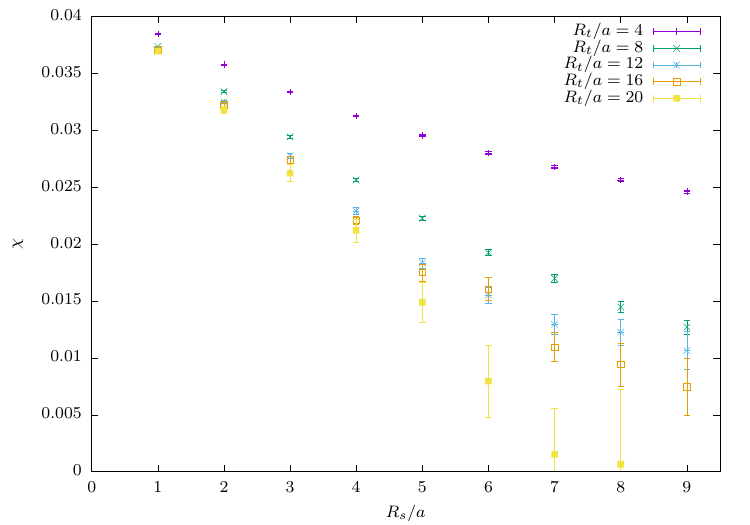}}
	\caption{Creutz ratios at two different masses at $\beta=35$. The smaller $\kappa$ (larger mass) is simulated on a $24\times 24$ lattice and the larger $\kappa$ on a $64\times 64$ lattice.}
\label{fig:stension1}
\end{figure}
One of our main goals is the investigation of confinement vs.\ screening scenario from the static quark-antiquark potential and the string tension in the limit of small fermion mass. With this data, we also wish the determine the behavior of the theory in the   chiral limit, as $m \rightarrow 0$. In this section we first focus on the case of $\Su{2}$ gauge group and add some remarks about the $N$ dependence in the end.

The static quark-antiquark potential $V$ is obtained from the scaling of the average Wilson loop ($W(\latR_s,\latR_t)$) with its temporal extend $R_t$ and spacial extend $R_s$ (in lattice units $\latR_t=R_t/a$ and $\latR_s=R_s/a$). It can be defined by 
\begin{align}
 aV(a\latR_s)=-\lim_{\latR_t\rightarrow\infty}\log(W(\latR_s,\latR_t))/\latR_t\; .
\end{align}
In two dimensional pure Yang-Mills theory, $V(R_s)$ scales linearly with $R_s$. The simplest ansatz for the scalar potential in two dimensions is therefore 
$V(R_s)=A+\sigma R_s$. The constant term ($A$) indicates a possible contribution with perimeter law.  
In the chiral limit of \AQCD, the theoretical expectation  for $\Su{2}$ is screening behaviour \cite{Gross:1995bp}:
$V(R_s)=c_1(1-\exp(-c_2R_s))$. For $\Su{\nc}$, as explained in Sec.~\ref{sec:con-sc}, if non-invertible symmetry is present, we expect a screening behaviour for all representations \cite{Komargodski:2020mxz}.  If $G_{\rm non-inv}$ is explicitly broken, we expect a confining potential  for  all representations, except for $N$-ality $k=0, \nc/2$ \cite{Cherman:2019hbq}.  
However, if the effects of 
$G_{\rm non-inv}$ breaking by the regularization are not relevant at the scales accessible in the simulation,  we again expect screening behavior in the massless fermion case. We argue that this is the case for our simulations with Wilson fermions. If we break $(\Z_2)_{\chi}$ explicitly by turning on a mass deformation, we always expect a confining behaviour. Consequences of these theoretical expectations are illustrated in Fig.~\ref{Potential}.

We consider several ways to obtain $V$ or its derivatives in this work. One simple approach is to use Creutz ratios. These correspond to discretised derivatives of  $\log(W(R_s,R_t))$ with respect to $R_s$ and $R_t$,
\begin{align}
 \chi(\latR_s,\latR_t)=-\log\left(\frac{W(\latR_s,\latR_t) W(\latR_s+1,\latR_t+1)}{W(\latR_s+1,\latR_t)W(\latR_s,\latR_t+1)}\right)\; .
\end{align}
If area law holds, $\chi$ should approach a plateau at large $R_s$ and $R_t$. The plateau value corresponds to the string tension.

We have tested this approach for several different parameters. At heavy fermion masses, $\chi$ is almost independent of $R_s$ and $R_t$, but in the low mass regime the string tension seem to decrease towards  zero at large distances, see Fig.~\ref{fig:stension1}. This may be either an indication of a very small string tension decreasing with $\sigma_k \sim m \lambda$ or potentially a screening behaviour.

Alternatively one can estimate the potential $V(R_s)$ in two different ways from the Wilson loop. It can be approximated from 
\begin{align}
 \hat{V}(\latR_s,\latR_t)=\log\left(\frac{W(\latR_s,\latR_t)}{W(\latR_s,\latR_t+1)}\right)\; ,
\end{align}
either by taking only the largest $\latR_t$ values into account (large $R_t$ method) or by a fit of the $\latR_t$ dependence in order to extrapolate to large $\latR_t$. The typical way to extrapolate the $\latR_t$ dependence is using the form $\hat{V}(\latR_s,\latR_t)=aV(a\latR_s)+\exp(-d \latR_t)$ in a fit to get $aV(a\latR_s)$ and $d$. We have implemented both methods, where, in case of the large $R_t$ method, we have averaged the values at the three largest $\latR_t$ values of the measurement. The measurement takes into account only $R_s$ and $R_t$ values up to half of the lattice size. In some cases we have restricted the data to smaller maximal $R_s$ and $R_t$ values if the signal over noise gets too small.

A further alternative is to estimate $V(R)$ from the correlator of two Polyakov lines at large $N_t$,
\begin{align}
 aV(a\latR_s)=-\log(\langle P(0)^\dag P(\latR_s) \rangle)/N_t\; .
\end{align}

\begin{figure}
	\centering
\subfloat[comparison of methods]{\includegraphics[width=0.48\textwidth]{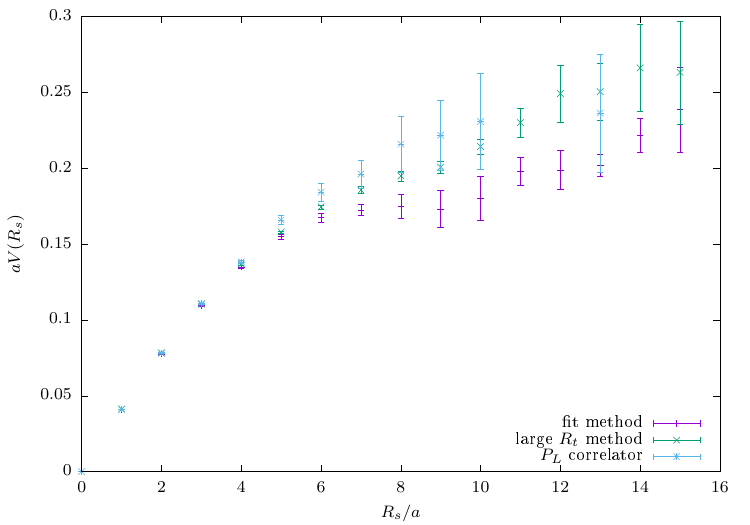}}
\subfloat[finite volume effects]{\includegraphics[width=0.48\textwidth]{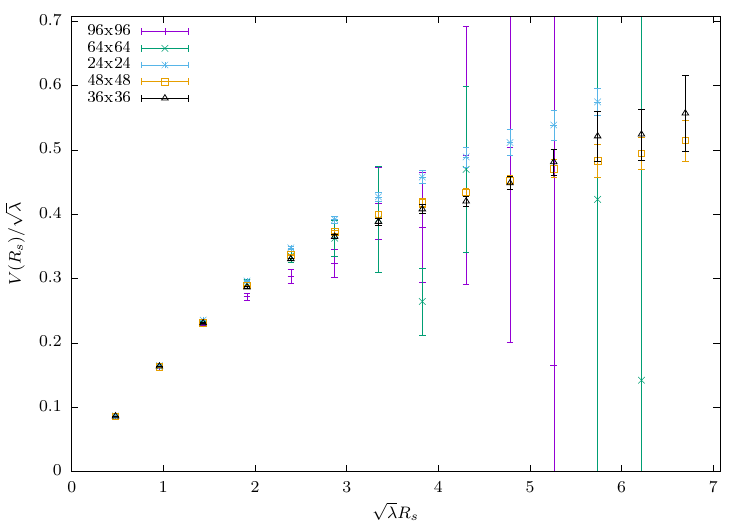}}
	\caption{Checks for systematic uncertainties of the static quark-antiquark potential at $\beta=35$, $\kappa=0.260$ on a $36\times 36$ lattice comparing different methods and volumes. The differences of the data represent the uncertainties related the estimation of the large $R_t$ limit and finite volume effects.}
\label{fig:stension2}
\end{figure}
We have applied all of the different methods in order to have consistent cross checks of the values and estimate systematic uncertainties. In addition we have also checked the volume dependence. As can be seen in Fig.~\ref{fig:stension2}, all of these tests lead to consistent results with some uncertainties at large $R_s$ and relevant finite size corrections only at the smallest volumes and lightest fermion masses.
Unless there are large uncertainties at larger $R_t$, we prefer to focus on the large $R_t$ method since it avoids further assumptions about the $R_t$ dependence. However, it might tend to overestimate $V(R_s)$ at large $R_s$.

\begin{figure}
	\centering
\includegraphics[width=0.48\textwidth]{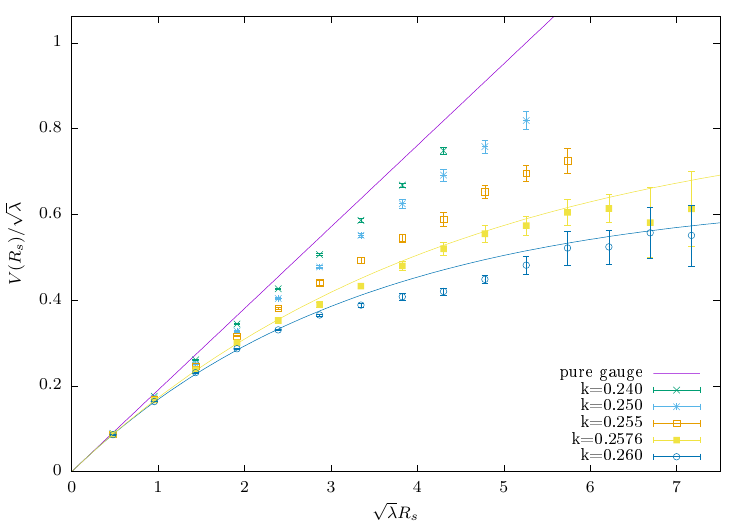}
	\caption{Mass dependence of static quark-antiquark potential at $\beta=35$ on a $36\times 36$ lattice. This fits of $V(R)$ at the smallest masses are according to $c_1(1-\exp(-c_2R))$ and the analytic prediction for the pure Yang-Mills case has been added. Results are obtained with the large $R_t$ method.}
\label{fig:stension3}
\end{figure}
\begin{figure}
	\centering
\subfloat[$V(R_s)$ linear fit]{\includegraphics[width=0.48\textwidth]{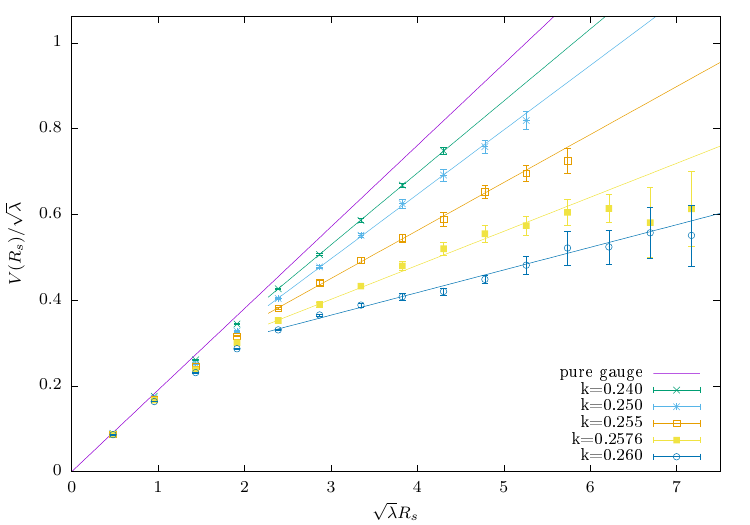}}
\subfloat[parameter linear fit]{\includegraphics[width=0.48\textwidth]{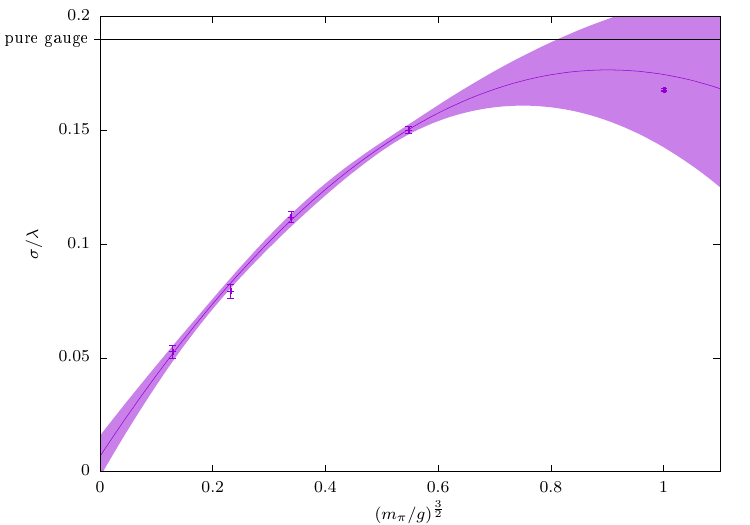}}
	\caption{Mass dependence of static quark-antiquark potential like in Fig.~\ref{fig:stension3}. (a) $V(R_s)$ at large $\sqrt{\lambda}R_s>2$ is fitted to the linear dependence $V(R_s)=A+\sigma R_s$. (b) The fit results are extrapolated to the chiral limit using a quadratic polynomial. Similar results are obtained for the gauge group \Su{3}, see Fig.~\ref{fig:stension3aSu3}. }
\label{fig:stension3a}
\end{figure}

The results for $V(R_s)$ confirm our findings obtained from the Creutz ratios. At large masses a linear raising potential with a string tension slightly smaller than in pure Yang-Mills theory is observed. Decreasing the fermion mass,  the slope in the  potential drops significantly. For a systematic determination of the long distance behaviour, compare Fig.~\ref{fig:stension3} and Fig.~\ref{fig:stension3a}. In the former, the potential is fitted to a  screening potential of the form $c_1(1-\exp(-c_2 R_s))$, while in the latter, it is  fitted to a linear potential at large $R_s$ to obtain the string tension. 
Theoretically, our expectation for small masses  is the one shown in Fig.~\ref{fig:stension3a}, while at exactly massless point (assuming $G_{\rm non-inv}$), we expect a screening potential. 
Indeed, 
the string tension tends to zero in the chiral limit Fig.~\ref{fig:stension3a}(b). 
Note, however, that the linear fit region might move towards larger values of $R_s$ in the chiral limit and the method becomes less reliable. At very small masses, we are not able to distinguish a small string tension from a screening potential.

\begin{figure}
	\centering
\subfloat[$m_\pi/g\approx 0.692$ ($\kappa=0.255$ at $\beta=35$)]{\includegraphics[width=0.48\textwidth]{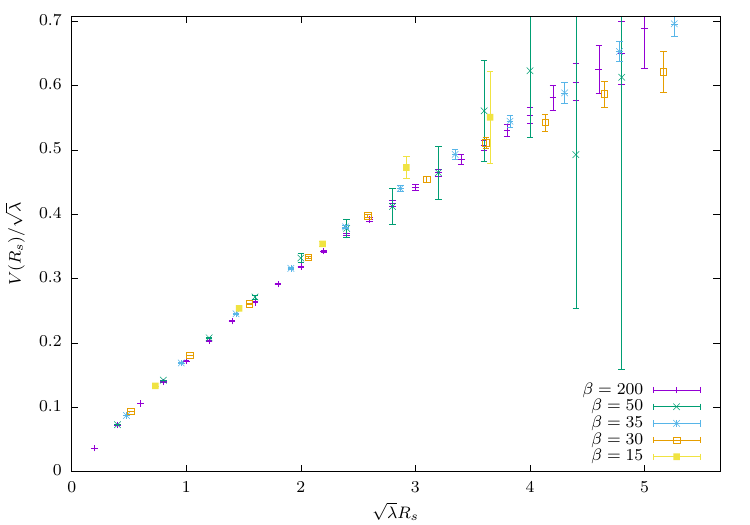}}
\subfloat[$m_\pi/g\approx 0.444$ ($\kappa=0.2576$ at $\beta=35$)]{\includegraphics[width=0.48\textwidth]{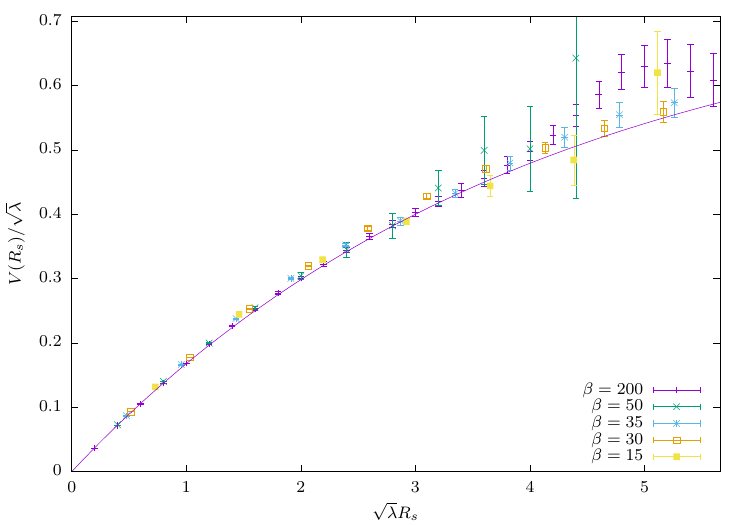}}
	\caption{Combined plot of $V(R)$ showing universal behaviour in units of the coupling $g$. The mass has been fixed to approximately the same value. Different lattice sizes have been combined in this plot: (a) $\beta=200,50$: $64\times 64$; $\beta=15,30$: $36\times 36$; $\beta=35$: $24\times 24$. (b) $\beta=200,50$: $64\times 64$; $\beta=15,30,35$: $36\times 36$. In (b) a fit to $c_1(1-\exp(-c_2R_s))$ has been added.}
\label{fig:stension4}
\end{figure}
\begin{figure}
	\centering
\subfloat[$\kappa=0.255$]{\includegraphics[width=0.48\textwidth]{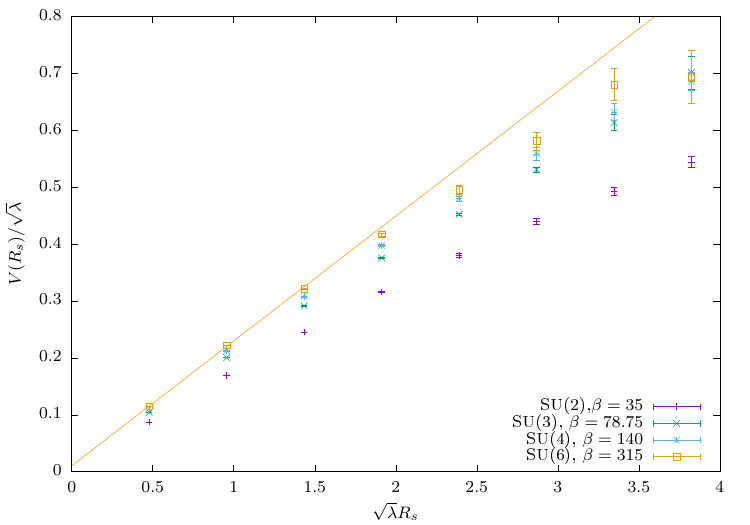}}
\subfloat[$\kappa=0.2576$]{\includegraphics[width=0.48\textwidth]{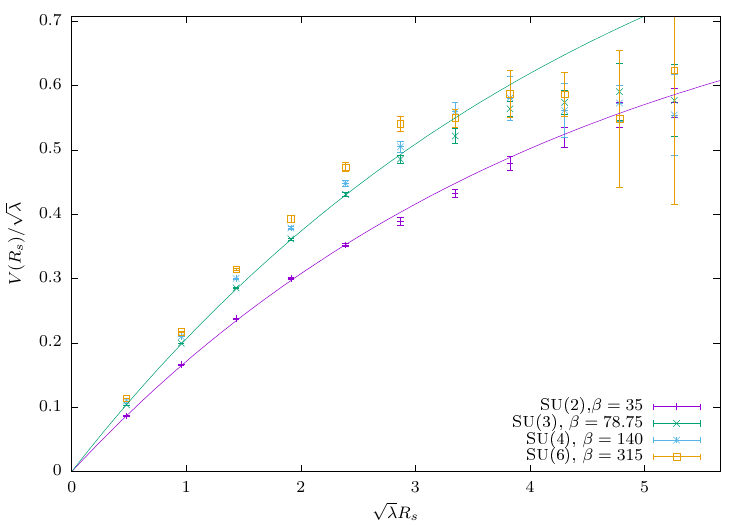}}
	\caption{$\nc$ dependence of $V(R_s)$ at fixed $\lambda$ and $\kappa$. In (a) a linear fit of $V(R_s)$ at $\nc=6$; in (b) a fit to $c_1(1-\exp(-c_2R_s))$ at $\nc=3$ and $2$ has been added.}
\label{fig:stension4SUN}
\end{figure}
The remaining step is the extrapolation of the results towards the continuum limit. This can be done by varying $\beta$ keeping $m_\pi/g$ fixed. This basically leads to consistent results and not much scaling towards the continuum limit is observed, see Fig.~\ref{fig:stension4}.

Towards larger $\nc$, the string tension increases at fixed $\lambda$ and $\kappa$, see Fig.~\ref{fig:stension4SUN}. The short distance $V(R_s)$ shows in general a larger slope than for $\nc=2$. However, at the larger distance behaviour at small masses,  our simulations  cannot distinguish screening from a confining potential with a parametrically small tension. 

Recall from theoretical discussion that the theory with exact $G_{\rm non-inv}$ symmetry is in the deconfined screening phase, and  the theory becomes confining if $G_{\rm non-inv}$  is explicitly broken 
by either mass operator ${\cal O}_{\chi}$ or by the ${\cal O}_{2}$ operator. The former also breaks the chiral symmetry, but the latter does not. Our current simulation is  not  at a stage to distinguish these two types of confinement. This stands as an important numerical problem.

\subsection{Lightest fermion and boson states}
\label{sec:light_states}
\begin{table}
    \centering
    \begin{tabular}{|c|c|c|c|c|c|}
    \hline
        $\beta$ & $\kappa$ & $m_\pi/\sqrt{\lambda}$ &  $m_{gg}^2\frac{\pi}{\lambda}$ & $am_{S}^2\frac{\pi}{\lambda}$ \\
         \hline
         35& 0.25& 0.6651(28)& 5.52(61)& 9.14(41)\\
         35& 0.255& 0.4859(50)& 3.64(58)& 7.52(87)\\
         35& 0.2576& 0.3727(65)& 3.4(1.1)& 7.10(97)\\
         15& 0.255& 0.6387(50)& 4.45(16)& 9.08(45)\\
         15& 0.2625& 0.4739(31)& 3.20(24)& 8.69(46)\\
        \hline
    \end{tabular}
    \caption{Bound states masses of the gluino-glue operator ($m_{gg}$) and the scalar meson ($m_S$).}
    \label{tab:masses}
\end{table}
A bosonic state can be formulated in terms of the mesonic scalar operator $S_\psi$ of \eqref{eq:op_meson}. On the lattice, this has to be measured from connected and disconnected fermion contributions similar to the mesons in four-dimensional $\mathcal{N}=1$ supersymmetric Yang-Mills theory. Consequently the same methods can be applied here, see \cite{Ali:2019gzj} for further details. In this first investigation, we have just applied a simple stochastic estimation of the disconnected contributions without further smearing or improvements of the operators.

In four-dimensional $\mathcal{N}=1$ supersymmetric Yang-Mills theory, the fermionic constituent of the supersymmetry bound state multiplet is described by a bound state of fermions and gluons (gluino-glue) in the following way
\begin{align}
O_{gg}=\sum_{\mu,\nu} \left[\gamma_\mu,\gamma_\nu\right]\Tr\left[ F_{\mu \nu} \lambda\right]\; .
\end{align}
A consistent projection operator onto a state for a correlator in time direction would require the indices $\mu$ and $\nu$ to run only over spacial indices. This is not possible in two dimensions but nevertheless the operator has been used in previous studies \cite{August:2018esp}. In this first study of the theory, we have applied this plain operator as well using a clover definition of the field strength $F_{\mu\nu}$ on the lattice. It is clear that link smearing can not be used to improve the operator and the first three smallest distances of the correlator have to be excluded.

All masses are extracted, as usual, from the exponential decay of the correlators at large distances. The window of the fit is restricted by the small signal to noise ratio at large distances.
Our first very rough estimates are shown in Tab.~\ref{tab:masses}. Note that masses for the lightest fermionic and bosonic states have been computed using Discretized Lightcone Quantization and Lattice Hamiltonean approach \cite{Dempsey:2022uie,Dempsey:2023fvm}.
The results are around $M^2\frac{\pi}{\lambda}=5.7$ for the lightest fermion and $M^2\frac{\pi}{\lambda}=10.8$ for the lightest boson. Our data are close to these values, but the systematic uncertainties are currently quite large.

\section{Conclusions and future directions}

This work aims  to initiate a systematic lattice study of \AQCD, with 
the hopes of  properly  understanding this  peculiar theory by both theoretical and numerical means.  
In our lattice formulation, we have used Wilson fermions, which require additive and multiplicative renormalization of the fermion mass.  
Hence the lattice formulation   does not respect the discrete chiral symmetry (which prevents  the mass term from being generated), as well as $G_{\rm non-inv}$ (which prevents the 4-fermion operator ${\cal O}_2$ and mass term from from being generated). 
To obtain chiral condensate in the massless limit, we used overlap fermions by reweighting. 
Our preliminary results indicate many agreements between theoretical understanding and numerical simulations. 
But we have not yet explored the realm of more difficult questions concerning the continuum massless theory with  $G_{\rm non-inv}$ symmetry. Below, we summarize our comparision of simulations with theory, and point some problems for future. 

With thermal compactification (apbc for fermions), we have numerically investigated the Polyakov loop $P$, its modulus $|P|$ and susceptibility.   At very high-temperatures, we find $\nc$ possible values, indicating a perturbative breaking of center symmetry.  However, for massive fermions, center-symmetry is restored non-perturbatively for finite $\nc$. Indeed, in the Polyakov loop scatter plot, we see that the simulation goes through all $\nc$ minima of holonomy potential, see Fig.~\ref{fig:nc6phase} for simulations, and Fig.~\ref{Wilson} for theory. The modulus of Polyakov loop goes from small values  at low temperature to approximately one at high-temperature. Furthermore, the susceptibility makes a peak around $T \sim (0.3-0.5)  \sqrt \lambda$, see Fig. \ref{fig:pltransition}.  However, it should be emphasized that this is not a phase transition, since $\langle P \rangle =0 $ on both sides. At  the large-$\nc$ limit, we provide evidence that  this cross-over becomes a sharp phase transition, see Fig.~\ref{fig:pltransitionNc}.
  It should also be pointed out that in a formulation which would respect $G_{\rm non-inv}$, the $N$-fold perturbative vacuum degeneracy must survive non-perturbatively because of mixed anomaly \eqref{algebra} even at finite-$\nc$.

We have investigated chiral condensate numerically for $\Su{2}$, see Fig.\ref{sfig:MChistCond}. It is consistent with theoretical expectations, and in particular, with the instanton result at high-temperature.  A rather beautiful result here is  Fig. \ref{fig:condpl}, 
Monte Carlo history of the correlation between the  Polyakov loop and condensate, indicating an instanton event in simulation and the fact that chiral condensate receives its dominant contribution from the instanton core. The  theoretical counterpart  is Fig.\ref{zeromode}, showing an instanton profile and a fermionic zero modes localized on the instanton core.  Fig.\ref{fig:condplcorrelation} is also evidence in this direction. We interpret these findings as implications of mixed anomaly between 
chiral symmetry and center-symmetry of massless theory once it is perturbed by a small mass term.  Usually, it is difficult to estimate the chiral condensate with Wilson fermions due to the additive renormalization. In two dimensions it is, however, possible to reweight the Wilson ensembles to the overlap operator. In this way, we have been able to determine the chiral condensate in the zero mass limit as shown in  Fig.~\ref{fig:fermioncond}.

We discussed in detail the current theoretical understanding of confinement vs. screening behavior of QCD(adj).  
The main theoretical result is shown in \eqref{current} and is plotted in Fig.~\ref{Potential}.  This understanding is based mainly on three works, \cite{Gross:1995bp, Cherman:2019hbq,  Komargodski:2020mxz}.
Clearly, this is a hard problem, and depending on whether  $G_{\rm non-inv}$ is explicitly broken or not, the answer changes.  Even the exactly massless QCD(adj) confines if 4-fermi operators are added. Our simulations show that the theory in the 
 massless limit tends towards  screening behavior.  As the fermion  mass gets smaller, the tension is reduced in a way proportional to fermion masses, see Fig.~\ref{fig:stension3a} and \ref{fig:stension3}.  
  We do not see an indication of 4-fermion induced confinement at the volumes we perform the simulations. Although the Wilson fermions do not respect non-invertible symmetry,   the coefficient of the induced 4-fermi operator ${\cal O}_2 $ given in \eqref{eq:def} must be rather small, 
  proportional to $ c_2 \sim r^2 a^2 g^2$ \cite{Cherman:2024onj}.  However, $c_2$ runs according to asymptotic freedom, and ultimately, it will become strong.  

There are further steps needed for a more complete analysis of the theory on the lattice. Obviously, the investigations of bound state masses require further improvements using other operators, like the baryonic ones discussed in \cite{Ali:2023zwl}, and better projections to the ground state using smearing. We also plan to extend the data at larger $\nc$, which is a thermodynamic limit, 
which may allow the study of genuine phase transitions, rather than cross-overs. 
Improvements of the lattice action might be helpful to get a complete picture, in particular using the overlap operator like applied in \cite{Bergner:2022hoo} in four dimensional adjoint QCD. The precise renormalization group scaling and the relevance of four fermion operators requires also further investigations in order to provide a more precise comparison to the results of \cite{Cherman:2019hbq,  Komargodski:2020mxz, Cherman:2024onj}.

Another direction we touched very little, but is important is the simulation of the theory with periodic boundary conditions. 
In this case, both theory Fig.~\ref{Wilson}(b)  and numerical simulations  Fig. \ref{fig:nc6phase}(b) indicate that a center-symmetric holonomy configuration \eqref{min-sym} is effective at  small-$L$. Thus, perturbative weak coupling analysis  (assuming $m=0$) indicates that there are $\nc-1$ types of massless fermions (living in the Cartan subalgebra)  at three level, unlike thermal compactification, where fermions are gapped-out perturbatively by acquiring a thermal mass.  Furthermore, periodically compactified (continuum) theory has at least four types of mixed anomalies between  the following symmetry pairs $ (\mathbb{Z}_N^{[1]},    (\mathbb{Z}_2)_\chi) $,  $ ( (\mathbb{Z}_2)_F,  (\mathbb{Z}_2)_\chi) $, 
$ ( (\mathbb{Z}_2)_C,  (\mathbb{Z}_2)_\chi) $,   and  $ (\mathbb{Z}_N^{[1]},   G_{\rm non-inv})  $ surviving compactification.
Thus, the vacuum structure of the compactified theory is partially explored in \cite{Cherman:2019hbq}
and  deserves further  studies  both theoretically (especially concerning the realization of non-invertible symmetry) and numerically.  
In our preliminary numerical work, we have already seen indications that this may be rather challenging due to zero modes or near-zero modes.

One more direction is performing simulations for different universes. This may have multiples utilities. One is, assuming that $G_{\rm non-inv}$ emerges at low-energies, then each universe must support exponentially large number of vacua, $\approx 2^{N-1}/N$. 
It may be interesting to investigate chiral condensates and  domain-wall (kink) properties  among these vacua, \eqref{eq:universe_vacua}.  The other is related to an order parameter for confinement. To see this, 
recall that standard partition function $Z(L) = \tr [e^{-L H}]$ sums over different universes without distinguishing them. The insertion of 1-form center symmetry generator ${\mathsf U_{\mathsf s}}^k (x)$ into the trace turns it into  a weighted sum over the universes, $Z_k(L) = \tr [e^{-L H} ({\mathsf U_{\mathsf s}})^k ] = \sum_{p=0}^{N-1}  e^{i \frac{2 \pi k p}{N}}   \widetilde Z_p $, where $\widetilde Z_p$
is the partition function for universe $p$, and ${\mathsf U_{\mathsf s}}^k (x)$ measures the charge associated with it (manifested in phase factor in the sum).  
Here, $k \in \Z_N$ is also equivalent to the insertion of $k$ units of 't Hooft flux, hence  $Z_k$ is partition function in 't Hooft flux sector $k$.   Therefore, the partition function for universe-$p$ is given by 
\begin{align}
   \widetilde Z_p = \sum_{p=0}^{N-1}  e^{i \frac{2 \pi k p}{N}}   Z_k
   \label{UPF}
\end{align}
\eqref{UPF} can also be interpreted as gauging the $\Z_N^{[1]}$ 1-form symmetry, which reduces the theory to $SU(N)/\Z_N$ theory.  Then, the theory can be probed by test charges (at infinity)  in the projective representation of the $SU(N)/\Z_N$, which corresponds to 
the  discrete  theta angle $\theta_p= \frac{2 \pi  p}{N}$ \cite{Witten:1978ka}, and  $\widetilde Z_p$ may also be viewed as partition function of $(SU(N)/\Z_N)_p$ theory. 
In two dimensions, the ratio   $\widetilde Z_p / \widetilde Z_0$ is nothing but the ratio of partition functions in the presence of  test-charges $\pm p$ at $\pm \infty$  vs. in its absence.  Hence, it is naturally an order parameter for confinement. If the ratio  scales as  $\exp(- \sigma A)$ where $A$ is the area of the Euclidean spacetime, then the theory is confined, otherwise it is deconfined.   Hence,  it may be useful to study lattice formulations directly projected into these universes.

\acknowledgments
We would like to thank Aleksey Cherman,  Theo Jacobson,  Yuya Tanizaki,  
Sahand Seifnashri, Zohar Komargodski, Mendel Nguyen, and Stefan D\"urr  for useful discussions. 
M.\"U. is supported by U.S. Department of Energy, Office of Science, Office of Nuclear Physics under Award Number DE-FG02-03ER41260. 
G.~B.\ is funded by the Deutsche Forschungsgemeinschaft (DFG) under Grant No.~432299911 and 431842497. Some computing time has been provided by the compute cluster ARA of the University of Jena and the compute cluster 
PALMA of the University of Münster.

\appendix
\section{Theory on the lattice}
\label{sec:lat}
The theory is formulated on a lattice of sizes $L_x=L_0=aN_x=aN_0$ and $L_t=L_1=aN_t=aN_1$ with lattice spacing $a$. In case of non-zero temperature studies, $L_t$ is considered as proportional to the inverse temperature, whereas $L_x$ should be large. If not further specified, the boundary conditions are periodic in all directions, except for the fermion fields, which have antiperiodic boundary conditions in the $t$-direction. In some cases, we have also investigated periodic fermion boundary conditions. The gauge theory is SU($\nc$). We mainly focus on fermions in the adjoint representation, but our program can handle also fundamental fermions.
\subsection{Lattice action and algorithms}
The gauge action is represented by a simple plaquette action. We are considering the Wilson fermion action in our main current simulations. In two dimensions, other more involved fermion formulations like fixed point and overlap fermions have been considered. As shown in \cite{Berruto:2002gn} for the 't Hooft model even a reweighting with the fermion determinant might be feasible\footnote{One difference to the 't Hooft model is, however, that the fermion determinant does not play a role in the large $\nc$ limit. This is different to the adjoint representation. Hence the scaling of the reweighting factors towards large $\nc$ is different.}. However, in these first simulations, the essential properties of the theory can be investigated already with a simpler fermion action. 
The Wilson-Dirac operator contains a mass parameter $m_0$ or, equivalently, the hopping parameter $\kappa=\frac{1}{2(m_0 a + 2)}$. Due to the breaking of chiral symmetry with this fermion action, the mass $m_0$ gets additive and multiplicative renormalization. The methods to handle the renormalization are well known in four dimensions and we explain some details in Sec.~\ref{sec:parameter_tuning}.

The simulations are based on the RHMC algorithm. We have used the program package of our studies of 4D supersymmetric Yang-Mills theory, which has been generalized to enable Monte-Carlo simulations in different dimensions. We have also developed a new independent python package based on tensorflow, capable of simulating these theories \cite{GitHubTf}.

We have tested and verified the code first in the case pure SU(2) Yang-Mills theory in two dimensions.  We have also cross checked our results with existing data for two dimensional QCD in fundamental and adjoint representation  \cite{Korcyl:2011tq,Korcyl:2012ub,Bibireata:2005sb}.

\subsection{Parameter tuning}
\label{sec:parameter_tuning}
The first important step is an identification of the relevant parameter space by choosing the mass parameter ($\kappa=\frac{1}{2(am_0+2)}$) and the gauge coupling ($\beta=\frac{2\nc}{ag^2}$). In addition the volume dependence has to be considered.

The bare gauge coupling $g$ in two dimensions has the units of mass. All other dimensionful quantities can be expressed in units of the bare gauge coupling. An important dimensionful quantity to quantify the relevant scale of the theory is the string tension. For a first estimate, it can be considered in the pure Yang-Mills limit. 
The scale of the string tension suggests large values of the gauge coupling if one assumes that the lattice size or lattice spacing in units of the string tension should be similar to common values in four dimensional lattice simulations. 

A further hint for the choice of the gauge coupling is the continuum extrapolation. The string tension in units of the gauge coupling can only be extrapolated to a continuum limit value if one excludes large values of $g$. This is related to the Gross-Witten transition \cite{Gross:1980he}. This transition separates a strong coupling and weak coupling phase at a value $\lambda=2$ of the 't Hooft  coupling $\lambda=g^2 \nc$. This constraint of the gauge coupling are naturally applied in studies of the  't Hooft model \cite{Berruto:2002gn}, but they should also be applied here. It implies a lower limit of the gauge coupling $\beta>\nc^2$. In order to have an feasible continuum extrapolation, even much larger values of $\beta$ should be considered.  In our SU(2) simulations, we have chosen $\beta=15,30,35,50,200$ and as a test considered also $\beta=3,6,10$. Towards larger SU($\nc$), we have tried to stay at constant 't Hooft couplings.

Concerning the values of the mass parameter, there are some difference with respect to the 't Hooft model due to the way the large $\nc$ limit is taken. Since in the 't Hooft model first the $\nc\rightarrow \infty$ and then the massless limit has to be considered, the mass in units of the gauge coupling has to be kept at a larger value. In our case, we are interested in the small mass limit even at a fixed $\nc$. 

\begin{figure}
	\centering
\subfloat[exponent fit\label{fig:fitmexponent}]{\includegraphics[width=0.48\textwidth]{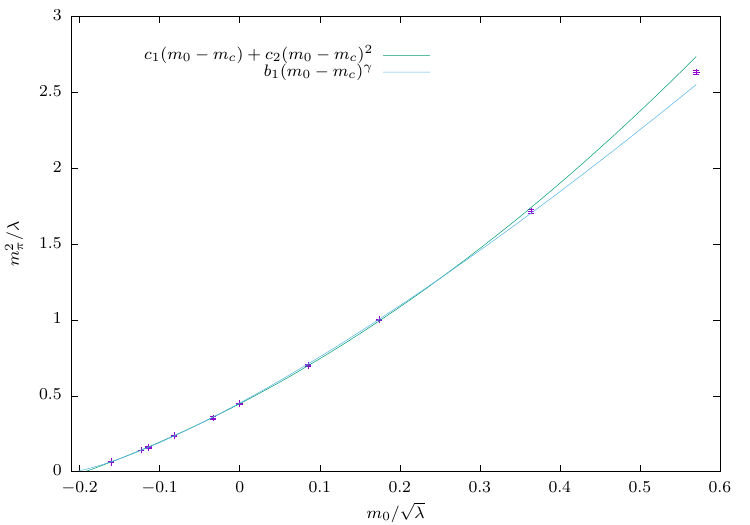}}
\subfloat[$\beta$ dependence\label{fig:fitmdepbeta}]{\includegraphics[width=0.48\textwidth]{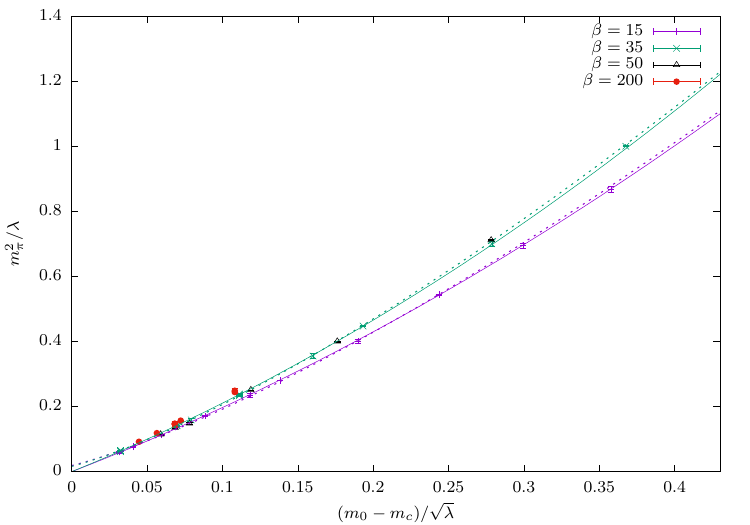}}
	\caption{Mass extrapolations in SU(2) \AQCD. Different volumes and gauge couplings are combined in this plot, excluding $L m_\pi<3$. (a) The data at $\beta=35$ is fitted according to a quadratic polynomial (fitting $m_c$, $c_1$, $c_2$) and with unknown exponent (fitting $m_c$, $b_1$, $\gamma$). The first fit result is $m_c/\sqrt{\lambda}=-0.19328(77)$, the second fit $m_c/\sqrt{\lambda}=-0.2084(14)$, $\gamma=1.311(13)$.
      (b) The data at $\beta=35$ and $\beta=15$ are fitted to a quadratic polynomial. The dashed line corresponds to an additional fit according to $b_1(x-m_c)^{4/3}$. The value of $m_c$ obtained in quadratic fit has been used to shift the data.}
	\label{fig:chirallimit}
\end{figure}
In the formulation with Wilson fermions, there is an additive and multiplicative remormalization of the fermion mass. Hence the tuning of the bare mass parameter ($m_0$ or $\kappa$) has to be done. This is usually based on signals for chiral symmetry, which is not a continuous symmetry in this model. The situation is quite similar to QCD with one fermion flavour in four dimensions and similar techniques can be applied here. The $a-\pi$ mass ($m_\pi$) can be an indication of the chiral limit (corresponding to parameters $m_c$, $\kappa_c$) and the mass tuning. 
It is derived from the correlator of $P$ in \eqref{eq:op_meson} by taking only connected fermion diagrams into account. It is not  related to a physical state, but it can be considered in a theory with a larger number of adjoint fermions. Our approach follows the techniques of partially quenched chiral perturbation theory used in the four dimensional counterpart.
The approximate dependence in four dimensions is provided by partially quenched chiral perturbation theory $m_\pi^2 \propto m$. This means $m_c$, the value of $m_0$ at which $m_\pi^2=0$, can be linearly extrapolated. 

In strong coupling large $\nc$ limit, the dependence  $m_\pi(m_0)$ is known analytically in four dimensions \cite{Kawamoto:1980fd}. The same analysis of weak and strong coupling limits provide a possible parameter range for our two dimensional model: $0.25\leq\kappa_c\leq 0.35$.
There might, however, be considerable differences to the four dimensional case. Expanding the strong coupling dependence in powers of $m_0$, the second order coefficient has the opposite sign compared to four dimensions, which indicates different corrections to the lowest order chiral extrapolation curve. Investigations of the 't Hooft model have even found a different functional dependence of $m_\pi$ on $m$. It has been observed that $m_\pi^2\propto m$ for U($\nc$), but $m_\pi\propto m^{2/3}$ for SU($\nc$) \cite{Hamer:1976bj,Berruto:2002gn}, which is also the expected dependence for the Schwinger model. 
We have tested different extrapolations with our simulation results as shown in Fig. \ref{fig:chirallimit}. A fit of the exponent in Fig.~\ref{fig:fitmexponent} yields a value closer to the dependence $m_\pi^2\propto m^{4/3}$, but a fit with linear dependence and quadratic corrections works equally well. At the current precision, it is therefore not possible to distinguish the two proposals for the mass dependence and the difference is almost negligible in the considered parameter range, see Fig.~\ref{fig:fitmdepbeta}. 
In addition we have also checked for possible finite volume effects. Considerable effects for $m_\pi$ are observed below $L m_\pi<3$.

We have considered also larger $\nc$ keeping the 't Hooft coupling $\lambda$ fixed. As an example, we have taken $\beta=35$ at SU(2) as a reference and obtained the results summarized in Tab.~\ref{tab:SUNmpi}. This implies that no separate mass tuning is required for different $\nc$ since the results at fixed $\lambda$ are consistent.

\begin{table}
    \centering
    \begin{tabular}{|c|c|c|c|c|}
    \hline
         & SU(2) & SU(3) & SU(4) & SU(6)\\
         \hline
        $\beta$ & 35 & 78.75 & 140 & 315\\
        \hline
        $am_\pi$ at $\kappa=0.255$ & 0.23239(34) & 0.23395(22) &0.23451(99) &0.23457(29)\\
        $am_\pi$ at $\kappa=0.2576$ & 0.18030(55) & 0.18155(28) &0.18309(25) &0.18285(43)\\
        \hline
    \end{tabular}
    \caption{Comparison of $m_\pi$ values for different $\nc$ keeping $\lambda$ constant.}
    \label{tab:SUNmpi}
\end{table}

\section{Further results}
We provide here some additional data to support our findings in the main part of the paper. 

\begin{figure}
	\centering
\includegraphics[width=0.48\textwidth]{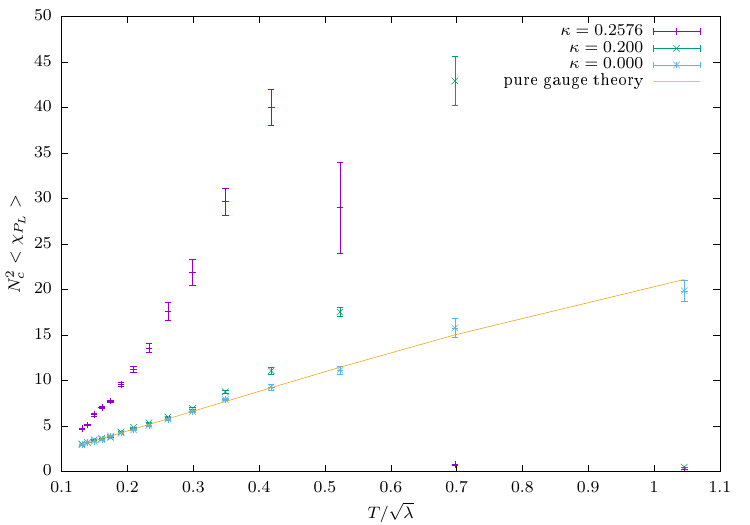}
	\caption{Extension of Fig.~\ref{fig:pltransitionMassDep} with the infinite mass limit, keeping the other parameters fixed. This time showing the susceptibility of the Polyakov loop (not of its modulus). The pure gauge line shows the analytic result. One lower and one heavy mass is added for comparison.}
\label{fig:infMassPhase}
\end{figure}
In order to show how the pure gauge limit is obtained from the mass dependence, we have added Fig.~\ref{fig:infMassPhase} as an extension of Fig.~\ref{fig:pltransitionMassDep}. 
The peak of the susceptibility moves towards higher temperatures as the mass increases if all other parameters are kept fix. It is expected to scale roughly like a fixed critical $mL$ if the mass is sufficiently heavy, see Eq.~\eqref{eq:massscale}. At temperatures below the peak, the susceptibility decreases with increasing masses, finally approaching the pure gauge limit. This limit can be analytically calculated using \eqref{eq:suscept} and  eq.14.30 in \cite{Wipf:2021mns}, if the susceptibility of the Polyakov loop is considered instead  of the susceptibility of its modulus as in the rest of the paper. The analytic prediction is consistent with our simulation data at $\kappa=0$, and is shown in Fig.~\ref{fig:infMassPhase}. 

The second additional investigation is whether the 
transition from a confining static potential with significant finite string tension towards screening or near zero string tension in massless limit can be observed  for $N \geq 3$. Therefore we repeated the investigation of  Fig.~\ref{fig:stension3a} for the gauge group \Su{3}. In pure gauge theory limit as well as at short distances in \AQCD, the slope of the static potential is larger than for \Su{2}. Therefore a fit of the large distance behaviour has to be done at a larger $\sqrt{\lambda}R_s$. The result is, however, quite consistent with the \Su{2} string tension: at zero mass limit, the string tension extrapolates to small values consistent with zero.  
\begin{figure}
	\centering
\subfloat[$V(R_s)$ linear fit]{\includegraphics[width=0.48\textwidth]{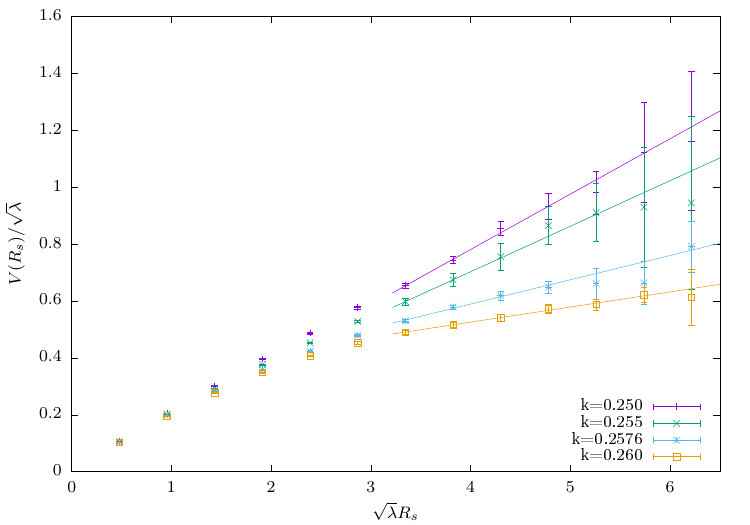}}
\subfloat[parameter linear fit]{\includegraphics[width=0.48\textwidth]{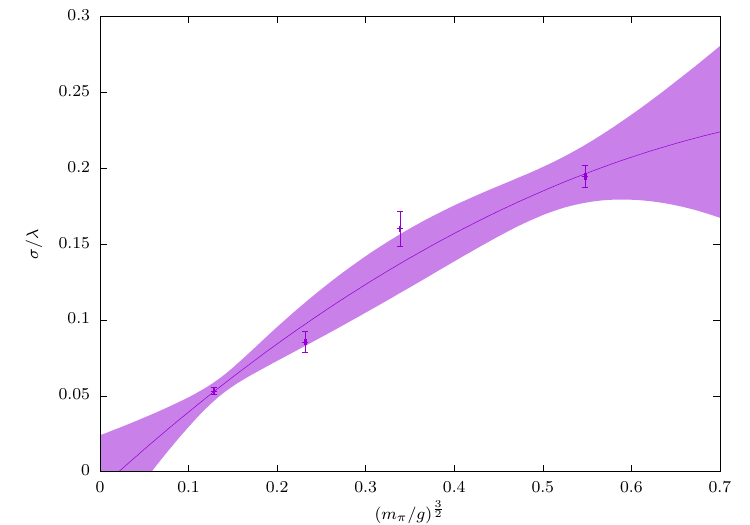}}
	\caption{Mass dependence of static quark-antiquark potential like in Fig.~\ref{fig:stension3a}, this time for the gauge group \Su{3}. (a) $V(R_s)$ at large $\sqrt{\lambda}R_s>3.2$ is fitted to the linear dependence $V(R_s)=A+\sigma R_s$. (b) The fit results are extrapolated to the chiral limit using a quadratic polynomial. }
\label{fig:stension3aSu3}
\end{figure}
\bibliographystyle{JHEP}
\bibliography{biblio}
\end{document}